\documentclass{aa} 

\usepackage[utf8]{inputenc}
\usepackage{graphicx}
\usepackage{booktabs}
\usepackage{txfonts}

\usepackage{xcolor}
\newcommand{\hrieuv}{HRI\textsubscript{EUV}}

\usepackage[colorlinks=true,linkcolor=magenta,citecolor=blue,filecolor=black,urlcolor=cyan,final=true]{hyperref}
\begin{document}

   \title{Fine-scale flare structures and their energetic implications from short-exposure extreme-ultraviolet imaging}
    \titlerunning{EUI flare} 
   \author{Laura A. Hayes
          \inst{\ref{dias}}
          \and
          Hannah Collier\inst{\ref{fhnw}, \ref{uot}, \ref{cita}}
          \and 
          S\"am Krucker\inst{\ref{fhnw}, \ref{ssl}}
          \and
          Daniel Ryan\inst{\ref{mssl}}
          \and
          David Berghmans\inst{\ref{rob}} 
          \and
          Emil Kraaikamp\inst{\ref{rob}}
          \and
          Muriel Stiefel\inst{\ref{fhnw},\ref{eth}}
          \and 
          Andrew Inglis\inst{\ref{gsfc}, \ref{cua}}
          }

   \institute{
            \label{dias}{Astronomy \& Astrophysics Section, Dublin Institute for Advanced Studies, Dublin, D02 XF86, Ireland.} \email{laura.hayes@dias.ie, lauraannhayes@gmail.com}
        \and 
            \label{fhnw}{University of Applied Sciences and Arts Northwestern Switzerland, Bahnhofstrasse 6, 5210 Windisch, Switzerland}
        \and
            \label{uot}{Department of Computer Science, University of Toronto, 40 St. George St., Toronto, ON, M5S 2E4, Canada}
        \and
            \label{cita}{Canadian Institute for Theoretical Astrophysics, 60 St. George Street, Toronto, ON M5S 3H8, Canada}
        \and 
            \label{eth}{ETH Zürich, Rämistrasse 101, 8092 Zürich Switzerland}
        \and
            \label{mssl}{University College London, Mullard Space Science Laboratory, Holmbury St Mary, Dorking, Surrey RH5 6NT UK}
        \and
             \label{ssl}{Space Sciences Laboratory, University of California, 7 Gauss Way, 94720 Berkeley, USA} 
        \and 
            \label{rob}{Solar-Terrestrial Centre of Excellence – SIDC, Royal Observatory of Belgium, Ringlaan -3- Av. Circulaire, 1180 Brussels, Belgium}
        \and 
            \label{gsfc}{Solar Physics Laboratory, Heliophysics Science Division, NASA Goddard Space Flight Center, Greenbelt, MD 20771, USA}
        \and
            \label{cua}{Physics Department, The Catholic University of America, Washington, DC 20064, USA}
}
   \date{Received xxx; accepted xxx}

 
  \abstract
   {To capture the brightest and most rapidly evolving phases of solar flares, the Major Flare Solar Orbiter Observing Plan (SOOP) employed a dedicated short-exposure mode for the High Resolution Imager (\hrieuv) of the Extreme Ultraviolet Imager (EUI), providing high-cadence, high spatial-resolution extreme-ultraviolet (EUV) imaging.}
   {We investigate the spatial and temporal organisation of compact flare emission and its implications for local energy deposition, using high-cadence short-exposure \hrieuv\ observations of the 2024 March 19 M2.1 flare.}
   {We combine the short-exposure (0.04~s) \hrieuv\ observations with hard X-ray (HXR) timing, imaging, and spectroscopic data from the Spectrometer Telescope for Imaging X-rays (STIX).  We characterise impulsive ribbon kernels and later loop strands, and compare footpoint areas measured with \hrieuv, the Atmospheric Imaging Assembly (AIA), and STIX to constrain the local energy flux carried by flare-accelerated electrons.}
   {The short-exposure observations reveal compact flare emission largely obscured by saturation in normal-exposure \hrieuv\ and AIA imaging. The spatially integrated \hrieuv\ emission evolves co-temporally with the STIX 22--45~keV hard X-ray emission, with no lag discernible beyond the 2~s \hrieuv\ sampling. The ribbons comprise repeatedly activated kernels with characteristic separations of ${\sim}1.4$--$1.7$~Mm, while the developing arcade shows a similar strand separation of ${\sim}1.3$~Mm, indicating that this ${\sim}1$--$2$~Mm spatial organisation persists into the newly formed flare loops. Measured kernel and strand widths of ${\sim}0.4$--$0.5$~Mm lie close to the instrumental resolution limit. The compact \hrieuv\ footpoint areas are approximately an order of magnitude smaller than those inferred from AIA or STIX, implying nominal local non-thermal energy fluxes on the order of $10^{11}$~erg~cm$^{-2}$~s$^{-1}$ at the HXR peaks. These estimates remain substantially higher across the tested intensity thresholds, although their absolute values depend on the area definition and spectral modelling}
   {The combination of high dynamic range, high spatial resolution, and high cadence in the \hrieuv\ short-exposure observations reveals a characteristic ${\sim}1$--$2$~Mm spatial organisation of the flare emission and substantially higher local non-thermal energy fluxes than inferred from conventional EUV or HXR source areas. These results demonstrate the value of flare-optimised EUV imaging for future solar flare observations.}

   \keywords{Sun: flares -- Sun: UV radiation -- Sun: X-rays, gamma rays -- instrumentation: high angular resolution}

   \maketitle
%
\section{Introduction}
\label{sec:intro}


Solar flares are explosive manifestations of magnetic energy release in the solar atmosphere, accompanied by rapid particle acceleration, plasma heating, and large-scale mass motions \citep{fletcher_2011}. In the standard picture, coronal magnetic reconnection releases stored free energy and accelerates particles, while energy transported along newly reconnected field lines drives heating and evaporation in the lower atmosphere, producing emission across hard X-ray (HXR), extreme ultraviolet (EUV), ultraviolet (UV), and optical wavelengths \citep{brown_1971, hudson_1972}. Within this framework, central questions remain open, in particular how the released energy is transported to the lower atmosphere and how it is organised in space and time. Addressing these questions requires observations that resolve the small-scale, fast-evolving ribbons, kernels, and loops in which energy deposition and the atmospheric response take place.

HXR and EUV observations are complementary probes of this process. HXR emission provides the most direct diagnostic of flare-accelerated electrons, locating the footpoints where non-thermal energy is deposited and constraining the electron spectrum and total power \citep{lin_2002, holman_2011, kontar_2011}. EUV and UV emission captures both the impulsive response of the lower atmosphere in flare ribbons \citep[e.g.][]{fletcher_2001, fletcher_2013} and the evolution of heated plasma in coronal loops, allowing the formation of ribbons, the growth of post-impulsive flare arcades, and the subsequent cooling to be tracked. Combining the two is therefore key to relating the drivers of energy release to the atmospheric response.

Flare ribbons provide a particularly direct view of this atmospheric response. They mark the lower-atmosphere footprints of newly reconnected field, but high-resolution observations show that they do not brighten as smooth, continuous fronts. Instead, they are composed of compact, localised intensity enhancements, hereafter called flare kernels (also described in the literature as beads, knots, or blobs). Observations with the Interface Region Imaging Spectrograph (IRIS) and ground-based facilities show that such kernels evolve rapidly and occur on scales of a few hundred kilometres, in some cases close to current instrumental resolution limits \citep{sharykin_2014, jing_2016, lorincik_2025, yadav_2025}. Bead-like brightenings and quasi-periodic ribbon substructure have been interpreted as possible chromospheric signatures of fragmented reconnection and tearing or plasmoid formation in the flare current sheet \citep{french_2021, wyper_2021, french_2025, dahlin_2025}. However, compact ribbon brightenings cannot be mapped uniquely to individual reconnection or plasmoid-formation events, as the observed ribbon morphology is also shaped by the magnetic geometry, radiative transfer, and variations in the local atmospheric response. Nevertheless, measuring the sizes, separations, and recurrence of compact kernels provides an indirect observational test of how flare energy release is organised in space and time.


Resolving these kernel scales is also central to flare energetics. The non-thermal energy flux delivered to the lower atmosphere, the electron power per unit footpoint area, is a key input parameter for radiative-hydrodynamic flare models \citep[e.g.][]{allred_2015, reep_2015}. If the compact footpoints are not resolved, whether in EUV, UV, or HXR imaging, the deposition area is overestimated and the inferred energy flux correspondingly underestimated. This flux influences the resulting chromospheric response, including whether the heating drives gentle evaporation, characterised by relatively slow upflows, or explosive evaporation, characterised by rapid upflows of hot plasma accompanied by downward-moving cooler material \citep{fisher_1985a, fisher_1985b, milligan_2006, polito_2016}. Although the transition between these regimes is not set by a single threshold but depends on the beam properties, heating duration, and atmospheric conditions \citep{reep_2015}, the adopted footpoint area remains a critical quantity for converting HXR-derived electron powers into physically meaningful local energy fluxes.

Measuring the impulsive EUV response at these scales is challenging. Full-disk imagers such as the Atmospheric Imaging Assembly \citep[AIA;][]{lemen_2012} on the Solar Dynamics Observatory (SDO) provide continuous multiwavelength coverage but are not optimised for the brightest, fastest-evolving flare cores. Saturation is a particular problem, and is not limited to the largest events: once the impulsive ribbon emission saturates, the brightest footpoint pixels are lost and the morphology, area, and temporal evolution of the compact kernels are distorted, while saturation in bright post-flare loops obscures the internal strand structure of the arcade. In addition, standard EUV cadences (e.g. 12~s for the AIA EUV channels) and spatial sampling from 1~au can undersample short-lived kernel-scale structure during the impulsive phase. Flare-optimised EUV observations therefore require sufficient dynamic range to avoid saturation, cadence fast enough to follow second-scale evolution, and sampling fine enough to resolve compact footpoints and loop strands.

Solar Orbiter \citep{mueller_2020} provides a unique opportunity to address these limitations, combining close-perihelion EUV imaging from the Extreme Ultraviolet Imager \citep[EUI;][]{rochus_2020} with HXR imaging spectroscopy from the Spectrometer Telescope for Imaging X-rays \citep[STIX;][]{krucker_2020} from the same vantage point. Dedicated EUI flare-observing modes use high-cadence, short-exposure imaging to mitigate saturation while revealing flare structure at the fine spatial scales accessible near perihelion. During the 2024 perihelion passages, EUI’s High Resolution Imager \hrieuv\ in 174~\AA\ operated in a dedicated short-exposure flare-watch mode as part of the Major Flare Solar Orbiter Observing Plan \citep[SOOP;][]{zouganelis_2020, ryan_2025}. This coordinated campaign was designed to capture flares at high spatial and temporal resolution with \hrieuv, STIX, Spectral Imaging of the Coronal Environment (SPICE), and Polarimetric and Helioseismic Image (PHI). This campaign has already enabled a series of complementary studies, revealing bead-like ribbon brightenings \citep{french_2025}, the small spatial scales and short heating timescales of transient EUV kernels \citep{collier_2026}, and evidence that different parts of flare ribbons can reflect different energy-transport mechanisms \citep{kerr_2026}. However, the relationship between the spatial scales of impulsive ribbon kernels and the subsequent flare loop arcade, and the implications of these scales for HXR-derived flare energetics, remain to be established.

Here, we analyse the M2.1 GOES-class flare of 2024 March 19, the first flare observed by \hrieuv\ in this short-exposure mode during the Major Flare campaign. The event was captured using alternating short (0.04~s) and normal (2~s) exposures, providing non-saturated 174~\AA\ imaging through the impulsive phase at high cadence. By combining these observations with STIX HXR timing, spectroscopy, and imaging, we (i) determine the spatial and temporal scales of the impulsive EUV kernels within the flare ribbons; (ii) test whether comparable spatial scales recur in the post-flare loop arcade; (iii) assess whether the compact EUV kernels provide a practical proxy for the electron-beam-impacted footpoint area when it is unresolved in HXR imaging; and (iv) quantify how this smaller area changes the inferred non-thermal energy flux and expected chromospheric response.


\section{Observations}
\label{sec:observations}

\subsection{Event overview and the Major Flare campaign}
\label{sec:obs:event}

\begin{figure}[t!]
    \centering
    \includegraphics[width=0.45\textwidth]{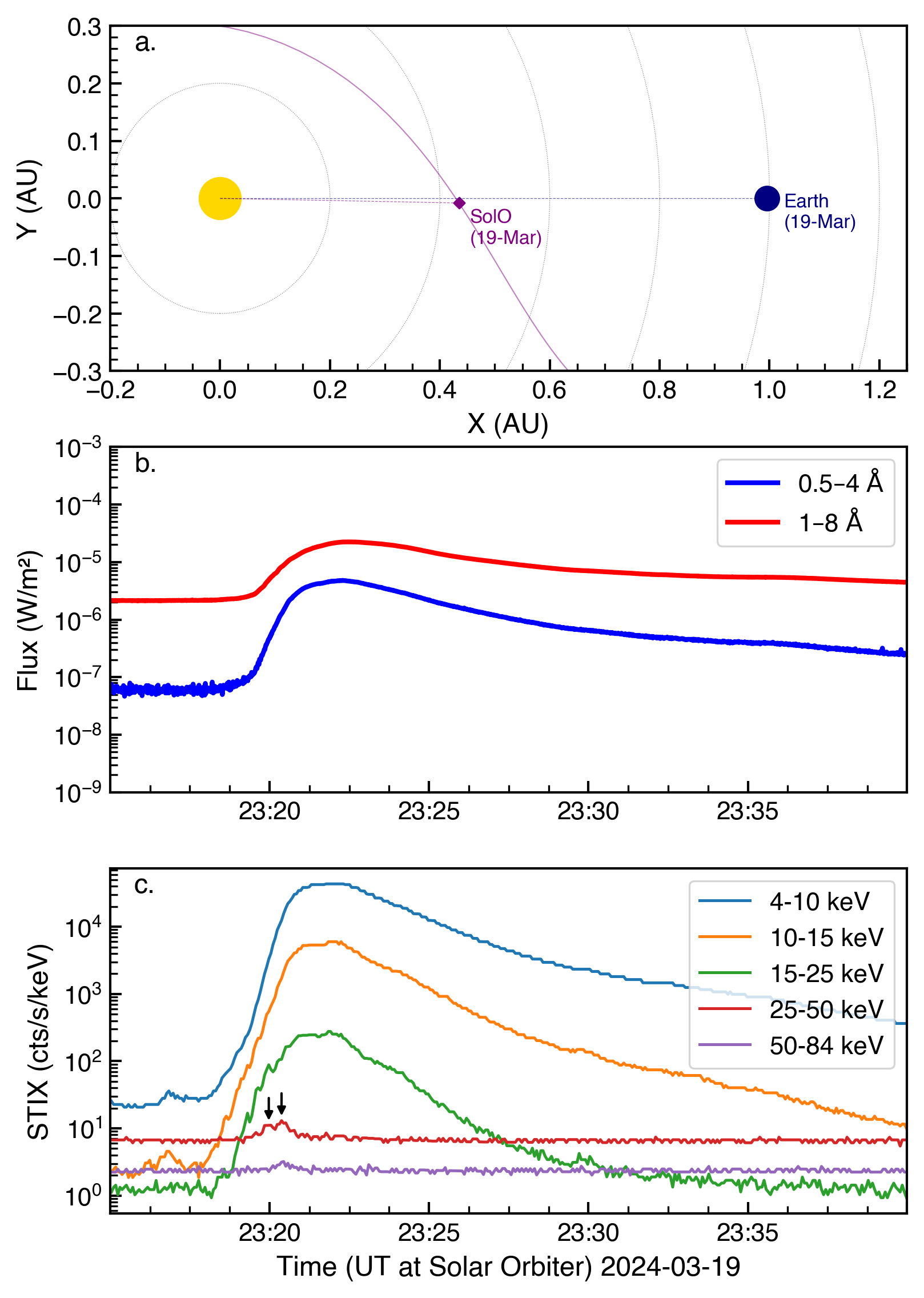}
    \caption{Overview of the 2024 March 19 M-class flare. \textit{(a)} Orbital geometry at the time of the flare: Solar Orbiter was located at ${\sim}0.43$~AU from the Sun and only ${\sim}1.3^\circ$ from the Sun--Earth line. \textit{(b)} GOES/XRS 0.5--4~\AA\ (blue) and 1--8~\AA\ (red) soft X-ray flux. \textit{(c)} STIX quicklook count rates in the 4--10, 10--15, 15--25, 25--50, and 50--84~keV bands. The impulsive hard X-ray burst (23:18--23:22~UT) precedes the soft X-ray peak, and the higher-energy bands display the second-timescale variability analysed in Sect.~\ref{sec:results:lightcurves}.}
    \label{fig:overview}
\end{figure}

\begin{figure*}
    \centering
    \includegraphics[width=\textwidth]{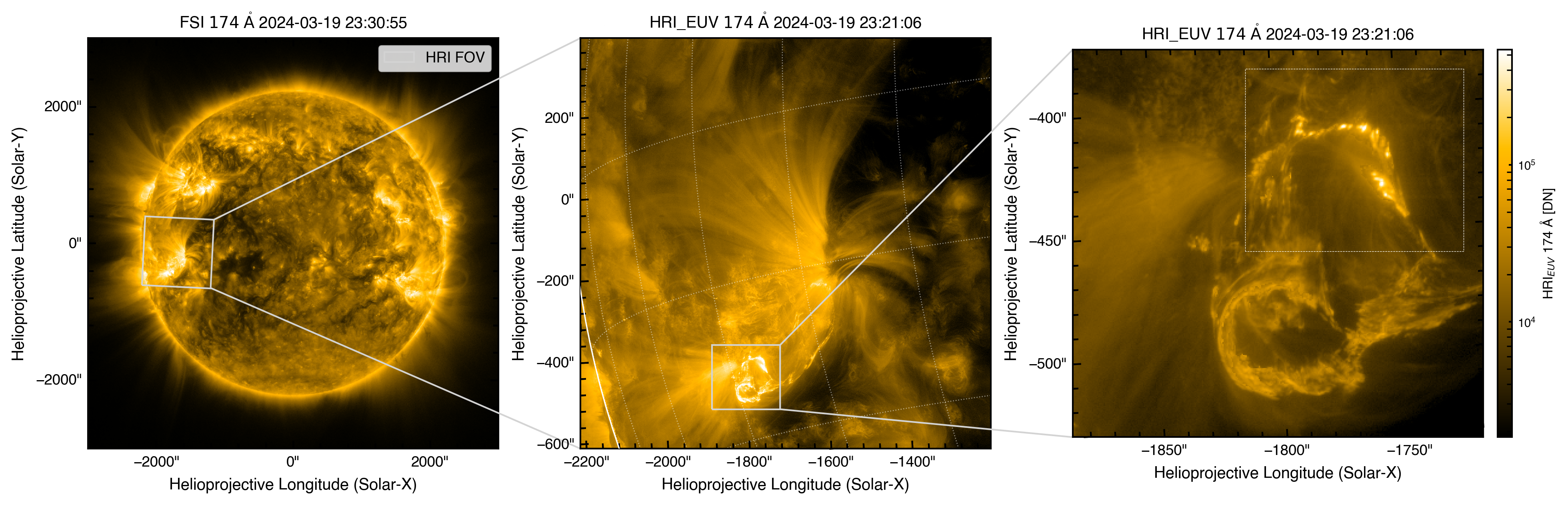}
    \caption{Flaring region in its full-disk context. \textit{(a)} EUI/FSI 174~\AA\ full-disk image at 23:30:55~UT with the \hrieuv\ field of view outlined (white box). \textit{(b)} Full \hrieuv\ 174~\AA\ frame at 23:21:06~UT; the box marks the flaring region near the eastern limb as seen from Solar Orbiter. \textit{(c)} Zoom onto the flaring region (combined normal- and short-exposure image; Sect.~\ref{sec:obs:eui}), showing the compact flare ribbons (dotted box) and the associated failed filament eruption (arrow).}
    \label{fig:fsi_hri}
\end{figure*}

The event analysed here is an M2.1 Geostationary Operational Environmental Satellite (GOES)-class flare that occurred on 2024 March 19 in NOAA active region 13615 which was observed by \hrieuv\ in the short-exposure flare-watch mode during the Major Flare SOOP \citep[see][for the campaign overview]{ryan_2025}. At the time of the observation Solar Orbiter was 0.43~au from the Sun and close to the Sun--Earth line (Fig.~\ref{fig:overview}~a). Unless stated otherwise, all times represent light-arrival time at Solar Orbiter, 279.9~s earlier than at Earth-orbiting assets. The flare was a compact two-ribbon event, with ribbons separated by ${\sim}24\arcsec$ as seen from Solar Orbiter (${\sim}7$~Mm on the Sun), accompanied by the failed eruption of a small filament channel whose footpoint brightened early in the impulsive phase (Sect.~\ref{sec:results:ribbon_lcs}). The GOES 1--8~\AA\ soft X-ray flux rises from ${\sim}$23:18~UT and peaks near 23:24~UT, while the STIX HXR emission defines an impulsive phase spanning 23:17:48--23:22:30~UT, with two dominant HXR peaks (marked by the arrows in Fig.~\ref{fig:overview}, and the shaded regions and arrows in Fig.~\ref{fig:lightcurves}~c). The \hrieuv\ field of view is overplotted on an EUI Full Sun Imager (FSI) image in Fig.~\ref{fig:fsi_hri}~a, b, and the combined short- and normal-exposure image in Fig.~\ref{fig:fsi_hri}~c provides a high-dynamic-range view of the compact ribbons and associated failed eruption.

\subsection{Solar Orbiter/EUI \hrieuv\ short-exposure sequence}
\label{sec:obs:eui}

The EUI observations are level-2 data products from EUI data release 6 \citep{eui_data_release_6}, consisting of \hrieuv\ 174~\AA\ images obtained during the Major Flare SOOP in a dedicated flare-watch mode designed to reduce saturation in bright flare emission while retaining the cadence needed to follow the rapid evolution of compact flaring structures. The sequence alternated six short-exposure images (effective exposure 0.04~s) with one normal-exposure image (2~s), giving effective cadences of 2~s and 16~s for the short- and normal-exposure sequences, respectively (see Figure~2 of \cite{ryan_2025}).

The \hrieuv\ images are $2048 \times 2048$ pixels with a scale of 0.492\arcsec~pixel$^{-1}$, corresponding to a projected ${\approx}0.155$~Mm~pixel$^{-1}$ at the Solar Orbiter distance during the observations, and a nominal two-pixel spatial resolution of ${\sim}0.31$~Mm. The passband is centred near 174~\AA\ (response ${\approx}171$--178~\AA, see Fig 21 and 22 from \cite{gissot}), and level-2 data are provided in data number per second (DN~s$^{-1}$) following the EUI calibration pipeline, including normalisation by integration time and pointing updates from nearby FSI limb-fitted images. Both exposure sequences were reprojected onto a common helioprojective grid using SunPy \citep{sunpy_2020}; the residual frame-to-frame jitter is below one pixel and does not affect the kernel and loop measurements, which are made from individual frames and local intensity profiles. Details of the co-registration, the combined images used for visualisation, and the onboard compression are given in Appendix~\ref{app:processing}.

\begin{figure*}
    \centering
    \includegraphics[width=0.9\textwidth]{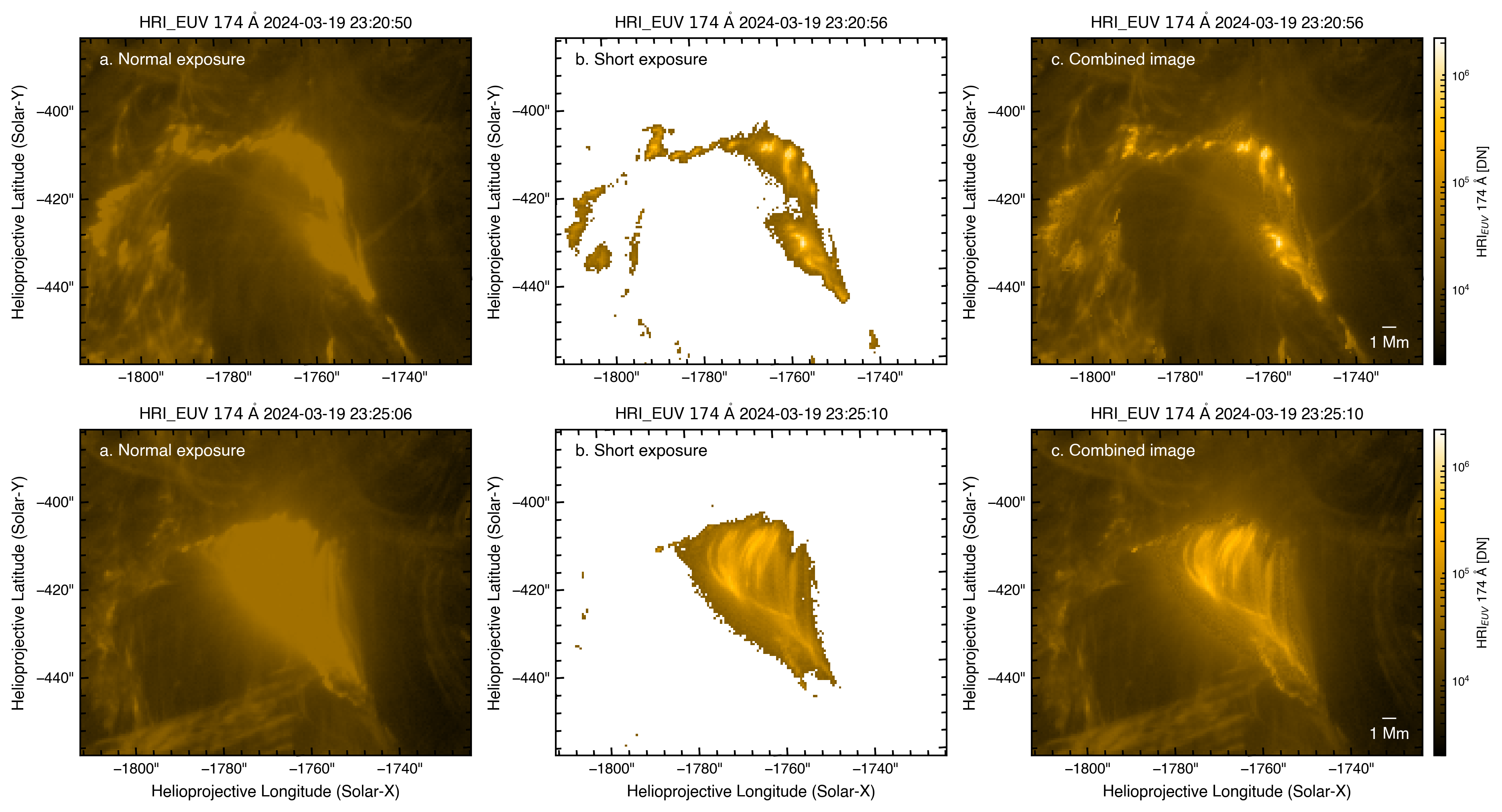}
    \caption{Comparison of \hrieuv\ normal- and short-exposure imaging during the impulsive phase (top row; 23:20:50--23:20:56~UT) and the early gradual phase (bottom row; 23:25:06--23:25:10~UT). \textit{(a)} Normal (2~s) exposures: the flaring core is overexposed, and saturates such that the internal structure of the ribbon is lost. \textit{(b)} Short (0.04~s) exposures: only pixels above the decided threshold are downlinked (faint pixels in the background 
    that were not downlinked have been put to white), and the compact ribbon kernels (top) and loop fine structure (bottom) are resolved without saturation. \textit{(c)} Combined image, in which the short-exposure data fill in the overexposed core of the normal-exposure frame. The scale bar marks 1~Mm.}
    \label{fig:fsi_hri_normal_short}
\end{figure*}

The diagnostic value of the alternating-exposure strategy is demonstrated in Fig.~\ref{fig:fsi_hri_normal_short}, where in the normal 2~s exposures the surrounding coronal structure is visible but the flare core is saturated and its internal ribbon or loop structure is lost, whereas in the near co-temporal 0.04~s short exposures the bright flare emission is recorded without saturation, revealing compact ribbon kernels during the impulsive phase and fine loop strands during the early gradual phase. All quantitative measurements of the compact flare emission are made from the short-exposure images.

\subsection{STIX, AIA, and GOES/XRS observations}
\label{sec:obs:stix_aia}
HXR diagnostics are provided by STIX, which observes 4--150~keV solar X-ray emission with spatially integrated spectroscopy and indirect Fourier imaging \citep{massa_2023}. For this observation, the STIX science pixel data have a temporal resolution of 0.5~s and are used throughout the analysis for the high-cadence HXR light curves, spectral fitting, and reconstruction of the non-thermal sources (Sect.~\ref{sec:results:energy}). The STIX quicklook count-rate light curves are used only to provide the broader event context in Fig.~\ref{fig:overview}. For comparison with a standard flare-monitoring EUV imager, we use SDO/AIA \citep{lemen_2012} level-1.5 imaging in 171~\AA\ (12~s cadence) and the 1600 and 1700~\AA\ UV channels (24~s cadence); the AIA EUV channels saturate severely during the impulsive phase (Fig.~\ref{fig:aia_eui_comparison}). We shifted the AIA times by the 279.9~s Earth--Solar Orbiter light-travel difference, and where a direct spatial comparison is required the AIA images are reprojected onto the Solar Orbiter viewpoint (Sect.~\ref{sec:results:energy}). The GOES-16 X-ray Sensor (XRS) 1--8~\AA\ flux provides the event classification and traces the bulk thermal heating.

\section{Results and discussion}
\label{sec:results}
We present and interpret the observations in four parts. We first establish the observational capability of the short-exposure data before examining the temporal evolution, fine structure, and energetics of the flare.


\subsection{Quantitative access to the flare core with short-exposure \hrieuv\ imaging}
\label{sec:results:dynamic_range}

\begin{figure*}
    \centering
    \includegraphics[width=0.8\textwidth]{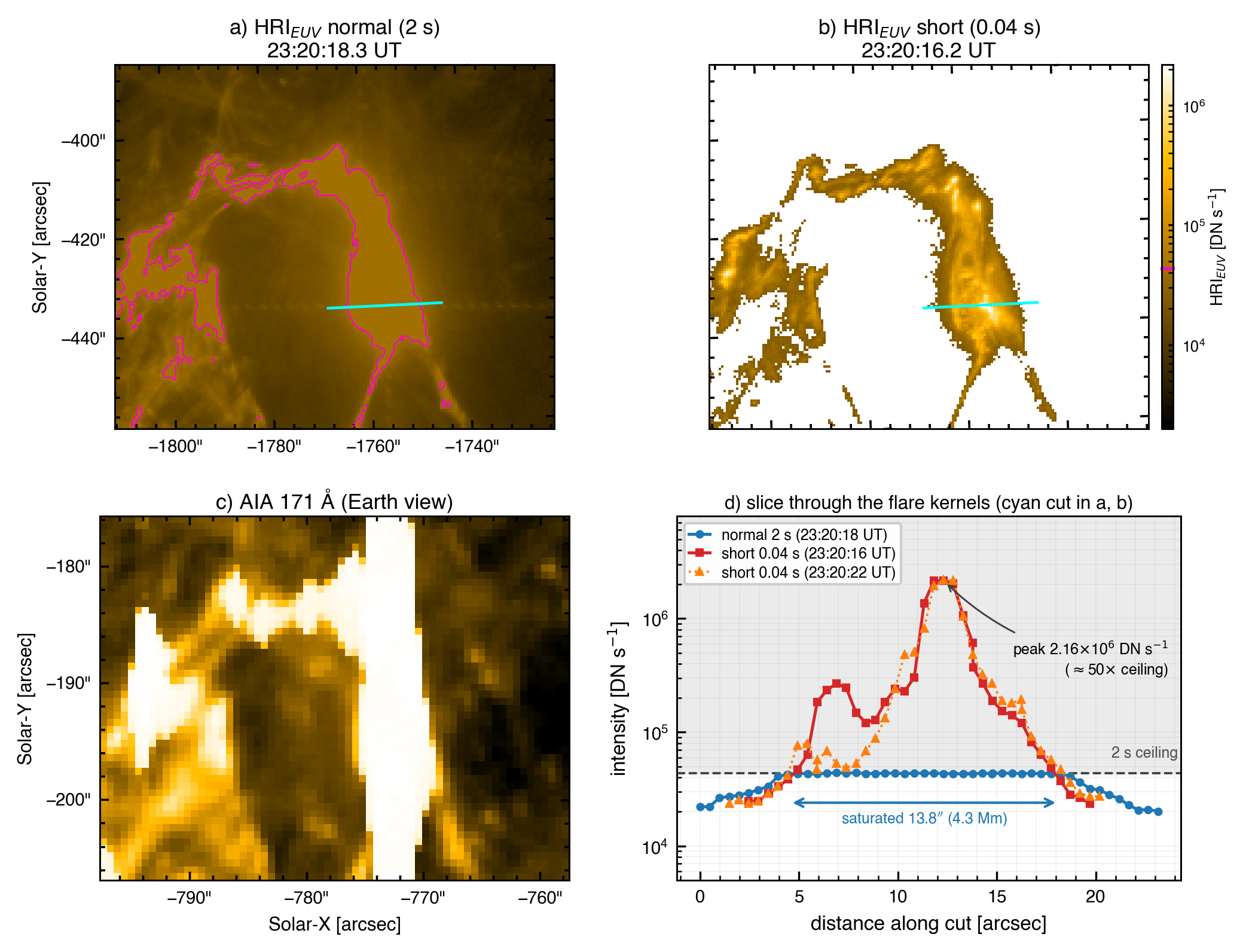}
    \caption{Dynamic range of the \hrieuv\ flare-watch mode during the impulsive peak. (a) Normal 2~s \hrieuv\ exposure, in which the flare ribbons saturate at the digital ceiling; the magenta contour marks the saturated region. (b) Nearly co-temporal 0.04~s short exposure on the same logarithmic DN~s$^{-1}$ scale. The magenta tick on the colour bar marks the 2~s saturation ceiling, and blank pixels in the short exposure image indicate data that were not telemetered. (c) Closest AIA 171~\AA\ image, shown at its native Earth-view resolution, illustrating saturation and blooming in the flare core. (d) Intensity along the cyan cut in panels (a) and (b). The normal exposure clips at $4.4\times10^{4}$~DN~s$^{-1}$.  The 0.04 s exposure increases the saturation ceiling in exposure-normalised units by a factor of 50, to approximately $2.2\times10^{6}$~DN~s$^{-1}$, greatly reducing the spatial extent of saturation and recovering the compact kernel profile; the brightest short-exposure pixels still reach this limit.}
    \label{fig:dynamic_range}
\end{figure*}

We first demonstrate the key observational advantage of the short-exposure \hrieuv\ sequence. Figure~\ref{fig:dynamic_range} compares the flare core during the impulsive phase in a normal 2~s \hrieuv\ exposure, a near-co-temporal 0.04~s short exposure, and AIA 171~\AA. Both the normal \hrieuv\ and the AIA 171~\AA\ images are strongly saturated in the flare core and do not preserve the compact substructure, whereas the short exposure records the same bright impulsive emission without saturation, revealing compact kernels embedded within the saturated region.

This difference is quantified by the cross-cut shown in Fig.~\ref{fig:dynamic_range}~d. Along the cut through the brightest kernel, the normal 2~s exposure reaches the saturation level over a distance of 13.8~\arcsec, corresponding to 4.3~Mm. The near-co-temporal short exposures instead recover the compact kernel profile over most of the region. Owing to the factor of 50 shorter integration time, the saturation ceiling in exposure-normalised units is increased by the same factor, from approximately $4.4\times10^{4}$ to $2.2\times10^{6}$~DN~s$^{-1}$. The brightest pixels in the short exposure can still reach this limit, but the spatial extent of saturation is greatly reduced, preserving the compact morphology that is lost in the normal exposure.

This substantial reduction in saturation is the main observational advantage of the flare-watch mode. It provides quantitative access to the morphology, areas, and temporal evolution of compact flare structures that are largely lost in the normal exposures, while retaining a 2~s cadence. It converts the flare core from a region in which information is lost through saturation into one in which intensities, areas, and light curves can be measured quantitatively at 2~s cadence. Desaturation algorithms can partially reconstruct saturated AIA cores \citep[e.g.][]{schwartz_2015}, and indeed \cite{guastavino_2026} used these same \hrieuv\ observations to validate Adaptive SE-DESAT, finding good agreement in the reconstructed morphology and relative flux evolution. Such methods, however, cannot recover information below AIA’s native spatial and temporal sampling. The EUI-STIX timing analysis, kernel and loop-strand statistics, and footpoint-area measurements presented below therefore rely on the combination of unsaturated imaging, high cadence, and fine spatial sampling provided by this observing mode. Its advantage is not simply improved image quality, but access to quantitative observables that are otherwise unavailable during the brightest phases of the flare. Short-exposure imaging with EUI's FSI first demonstrated this principle \citep{collier_2024_eui} but \hrieuv\ extends this capability to substantially finer spatial scales while revealing both the temporal evolution of the kernels and the fine structure of the developing arcade, complementing normal-exposure \hrieuv\ flare studies at perihelion \citep[e.g.][]{chitta_2026}. By retaining the surrounding coronal context in the normal exposures while measuring the bright flare core in the short exposures, the alternating-exposure strategy offers a practical template for flare-optimised EUV observations within realistic telemetry constraints. Additional \hrieuv-AIA comparisons are shown in Appendix~\ref{app:aia_comparison}.

\subsection{Co-temporal evolution of the EUV kernels and HXRs}
\label{sec:results:lightcurves}

\begin{figure}
    \centering
    \includegraphics[width=0.95\columnwidth]{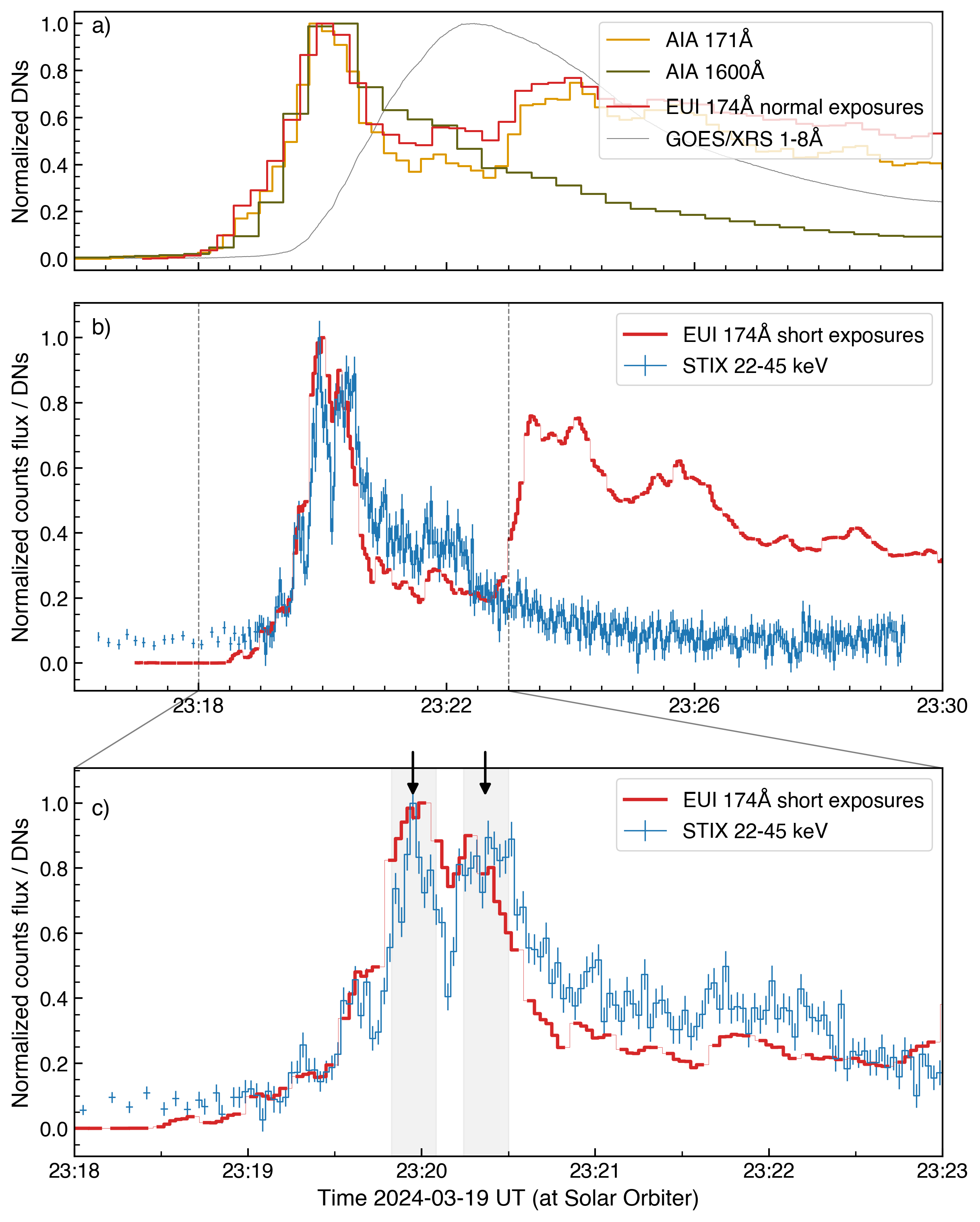}
    \caption{{Multi-instrument light curves through the impulsive and early gradual phases. All times are in the Solar Orbiter frame. \textit{(a)} Normalised AIA~171~\AA, AIA~1600~\AA, \hrieuv\ 174~\AA\ normal-exposure, and GOES/XRS 1--8~\AA\ light curves. \textit{(b)} Normalised \hrieuv\ 174~\AA\ short-exposure flux (red) and livetime-corrected STIX 22--45~keV count rate (blue, with error bars). The short-exposure series is interrupted every seventh frame by an interleaved normal exposure, these gaps are bridged by thin connecting lines and contain no short-exposure data. Dashed vertical lines mark the interval expanded in panel~(c). \textit{(c)} Expanded view of the impulsive phase, showing the close correspondence between the second-scale \hrieuv\ and HXR variability. Grey shaded intervals mark the STIX spectral-fit windows for Peaks~1 and 2 used in Sect.~\ref{sec:results:energy}.}}
    \label{fig:lightcurves}
\end{figure}

Here we focus on the relationship between the field-of-view (FOV) integrated, high-cadence \hrieuv\ short-exposure emission and the non-thermal HXR emission during the impulsive phase, as shown in Fig.~\ref{fig:lightcurves}. The \hrieuv\ 174~\AA\ emission contains two physically distinct components. During the impulsive phase, rapid heating of the lower atmosphere produces bright, compact ribbon kernels that are expected to evolve closely with the HXR emission from flare-accelerated electrons \citep{brown_1971, hudson_1972, kontar_2011}. Later, evaporated plasma fills the newly reconnected loops and cools through the \hrieuv\ temperature response, producing a broader 174~\AA\ gradual-phase enhancement that is not directly associated with ongoing non-thermal electron precipitation. Both components appear in Fig.~\ref{fig:lightcurves}~a where the AIA 1600~\AA\ channel is predominantly impulsive, the AIA 171~\AA\ and \hrieuv\ normal-exposure 174~\AA\ light curves show both an impulsive rise and a later gradual-phase enhancement, and the GOES soft X-ray flux peaks after the impulsive phase. Saturation in the flare core compromises the spatial information in these integrated light curves, however, preventing the compact impulsive structure from being isolated.

The high-cadence short-exposure sequence isolates the impulsive flare core without saturation. As shown in Fig.~\ref{fig:lightcurves}~b, the spatially integrated \hrieuv\ short-exposure flux rises rapidly with the livetime-corrected STIX 22--45~keV emission, peaks during the HXR burst, and then decays as the non-thermal HXR flux weakens. From ${\sim}$23:22:30~UT onward, the 174~\AA\ signal rises again while the STIX 22--45~keV flux continues to decay, marking the transition from impulsive ribbon emission to gradual-phase loop emission. The zoomed in view (Fig.~\ref{fig:lightcurves}~c) shows that individual seconds-scale bursts in the \hrieuv\ flux coincide with bursts in the STIX HXR emission.

We quantify this relationship over the impulsive interval 23:18:30-23:22:00~UT, using the native \hrieuv\ short-exposure sampling and the STIX 22--45 keV light curve. The raw light curves have a Pearson correlation coefficient of $r=0.90$, reflecting their similar overall evolution during the impulsive phase. To test whether this correspondence is driven only by the shared flare envelope, we subtract a 21~s running mean from each light curve and compare the resulting shorter-timescale residuals. These remain correlated at $r=0.63$, with the cross-correlation consistent with zero lag. The rapid EUV and HXR variability is therefore co-temporal within the temporal resolution of the observations. The correspondence weakens after the impulsive phase as cooling-loop emission becomes increasingly important, so we use the compact kernels as tracers of the impulsive lower-atmosphere response only.

\begin{figure*}
    \centering
    \includegraphics[width=\textwidth]{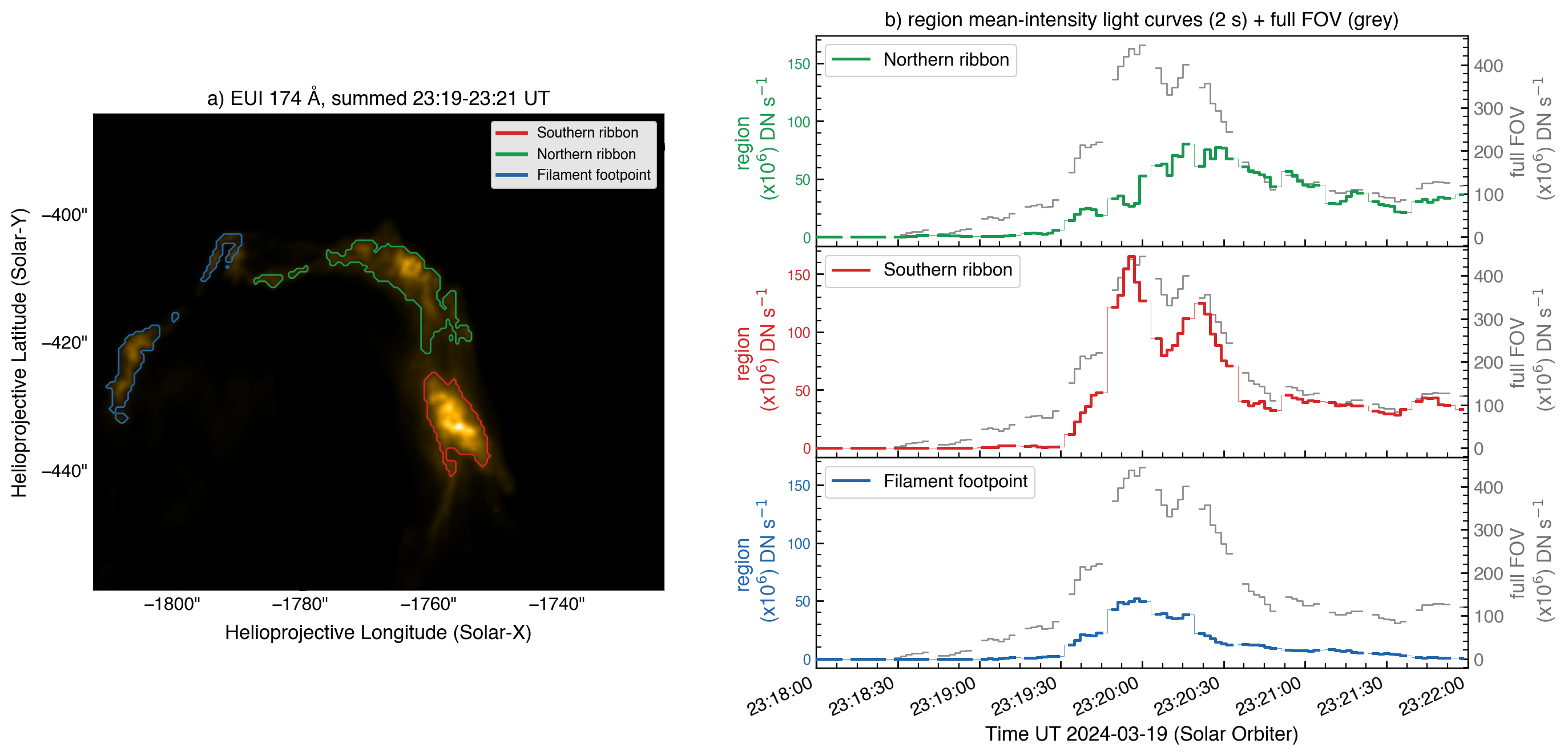}
    \caption{Region-resolved \hrieuv\ 174~\AA\ light curves during the impulsive phase. \textit{(a)} \hrieuv\ 174~\AA\ image summed over 23:19--23:21~UT (Solar Orbiter time), with contours outlining the three regions of interest: the southern ribbon (red), the northern ribbon (green), and the filament channel footpoint (blue). \textit{(b)} Mean-intensity light curves at 2~s cadence for each region (from top to bottom: northern ribbon, southern ribbon, filament footpoint; coloured curves, left axes), each shown against the full-FOV mean-intensity light curve (grey, right axes) for reference. The southern ribbon is characterised by two dominant double impulsive peaks near 23:20:20~UT followed by a smooth decay, while the northern ribbon shows a more fragmented, multi-peaked profile sustained across the impulsive phase. The filament footpoint brightens at a lower intensity than the two principal ribbons. All times are in the Solar Orbiter frame.}
    \label{fig:timeseries_ribbons}
\end{figure*}

We further exploit the spatially resolved short-exposure images to examine how the impulsive emission evolves across the flare\label{sec:results:ribbon_lcs}. From the short-exposure image summed over 23:19--23:21~UT (Fig.~\ref{fig:timeseries_ribbons}~a) we define three polygonal regions: the northern ribbon (P1), the southern ribbon (P2), and a filament-footpoint region (P3), and extract their 2~s-cadence mean-intensity light curves (Fig.~\ref{fig:timeseries_ribbons}~b).

The three regions show distinct temporal behaviour, demonstrating that the impulsive energy deposition is spatially inhomogeneous. The southern ribbon is dominated by a compact, rapidly rising burst near the main impulsive peak, followed by a secondary enhancement and decay on a timescale of about one minute. The northern ribbon evolves more gradually and shows a broader, multi-peaked profile sustained across much of the impulsive phase. The filament-footpoint region shows a lower-intensity, more compact enhancement than the two principal ribbons, consistent with its association with the early evolution of the failed filament eruption. This structure, largely hidden in spatially integrated or saturated observations, shows that the flare does not brighten as a single coherent ribbon-wide front, consistent with the kernel-level analysis in Sect.~\ref{sec:results:kernels}.

Taken together, these timing relationships support interpreting the compact 174~\AA\ kernels as closely associated with the impulsive energy-release process. Compact UV and EUV footpoints have long been read as tracers of impulsive energy deposition and the chromospheric endpoints of newly reconnected field \citep{fletcher_2009}, and the present observations extend this by showing that the impulsive EUV response can be measured quantitatively, without saturation, throughout the brightest phase, consistent with rapid compact footpoints seen with TRACE \citep{fletcher_2004}, the resolved atmospheric response and chromospheric evaporation revealed by IRIS observations \citep{tian_2015, graham_2015, battaglia_2015, polito_2016, lorincik_2025} as well as recent \hrieuv\ kernel statistics \citep{collier_2026}. We emphasise, however, that the EUV kernels do not image the electron beam directly, but trace the atmospheric response at compact footpoints that is co-temporal, within the cadence, with the HXR-producing electron population. Two caveats follow. First, the 174 Å passband can respond to plasma heated by mechanisms other than electron beams, including thermal conduction and potentially the dissipation of downward-propagating Alfvénic waves \citep{reep_russell_2016, reep_2018}, so an EUV kernel need not be exclusively beam-heated. For example, coordinated SPICE, EUI, and STIX observations show that different ribbon segments can reflect different energy-transport mechanisms \citep{kerr_2026}. Second, the association with STIX is established from the temporal correspondence of the spatially integrated EUV and HXR emission, rather than through a one-to-one identification of individual EUV kernels with HXR sources. The separately evolving filament-channel footpoint illustrates this complexity, similar to the distinct HXR-emitting filament anchor points reported by \citet{stiefel_2023}.

\subsection{Fine structure of ribbon kernels and flare loop strands}
\label{sec:results:kernels}

\begin{figure}[!t]
    \centering
    \includegraphics[width=0.9\columnwidth]{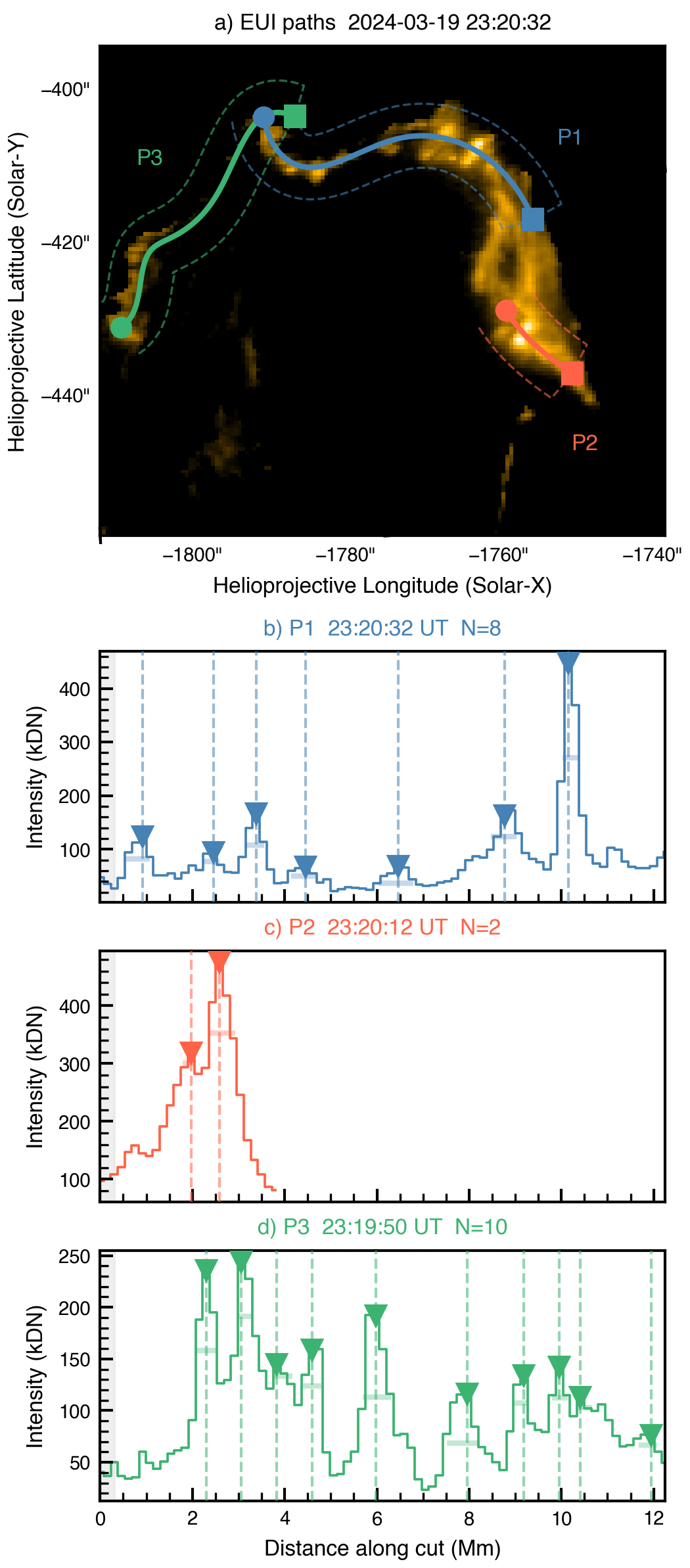}
  \caption{Impulsive-phase ribbon fine structure. (a) \hrieuv\ 174~\AA\ image at 23:20:32~UT, with curved paths tracing the North ribbon (P1, blue), South ribbon (P2, red), and filament footpoint (P3, green); dashed lines mark the 17-pixel-wide averaging slit, and the circle and square denote the start and end of each path. (b)--(d) Intensity profiles along P1, P2, and P3 (at 23:20:32, 23:20:12, and 23:19:50~UT). Kernels, identified after Savitzky--Golay detrending as peaks of fractional prominence $\geq0.08$, are marked by triangles and dashed lines; horizontal bars show the half-prominence width. The grey band marks the two-pixel resolution limit ($0.31$~Mm), and $N$ is the number of kernels detected.}
    \label{fig:imp_slice}
\end{figure}

We next quantify the fine structure within the impulsive ribbons using the non-saturated \hrieuv\ short-exposure images. Along curved spine paths tracing the northern ribbon (P1), the southern ribbon (P2), and the filament footpoint (P3), the same regions as in Sect.~\ref{sec:results:ribbon_lcs}, we extract one-dimensional intensity profiles averaged over a 17-pixel-wide (${\approx}2.6$~Mm) slit oriented perpendicular to the local path. This averaging provides a representative along-ribbon profile while smoothing some finer structure perpendicular to the ribbon, and similarly, the 2~s cadence provides high-cadence sampling of the kernel evolution, although still faster changes between frames may not be captured. These paths are shown in Fig~\ref{fig:imp_slice}~a. Each profile is detrended with a Savitzky--Golay filter (31-pixel, ${\approx}4.8$~Mm window, second-order polynomial) to subtract the smooth ribbon envelope (this window is an order of magnitude larger than the kernel widths and about three times their typical separation, so the fine structure is passed essentially unattenuated while the several-Mm envelope is removed), and compact kernels are identified as peaks of fractional prominence $\geq0.08$, i.e. by local contrast rather than absolute intensity. The separation between adjacent kernels is the peak-to-peak distance along the path, and the kernel width is the full width at half prominence (which captures the local extent of each brightening above its surrounding inter-kernel background), measured on the native pixel-scale profiles (the same width definition is used below for the loop strands). 

Figure~\ref{fig:imp_slice} illustrates the procedure for representative frames. The extracted profiles show that the impulsive ribbons are not smooth structures but are composed of discrete, localised bright kernels in all three regions, with the northern ribbon containing multiple kernels along an extended path, the southern ribbon dominated by a few bright kernels, and the filament footpoint region shows several short-lived brightenings during the early impulsive phase. The distributions of kernel separations and widths, pooled over the impulsive-phase frames, are summarised in Table~\ref{tab:scales} (full distributions in Fig.~\ref{fig:stats_imp_kernels}, Appendix~\ref{app:distributions}). Despite their different temporal behaviour and morphology, all three regions show characteristic kernel separations of order 1--2~Mm. The kernel widths are typically ${\sim}0.4$--$0.5$~Mm, only a few pixels across. Although most measured widths exceed the two-pixel resolution limit of 0.31~Mm, they lie close to this limit and should therefore be interpreted as upper limits, with the true distribution possibly extending to smaller scales.

The kernel locations are also temporally persistent rather than randomly distributed. To quantify this we followed the along-path positions of all detections from frame to frame. For each detected kernel we asked whether another detection appears within the two-pixel resolution limit in the next $2$~s frame. Across the three paths, $75$--$83\%$ of detections satisfy this criterion, compared with ${\sim}30\%$ for randomised kernel positions. The brightenings therefore reactivate repeatedly at a limited number of preferred locations along each path, rather than appearing at independent
positions in each frame. This is visible in Fig.~\ref{fig:spacetime} as near-horizontal tracks persisting over tens of seconds, separated by intervals in which the same position is dark. The behaviour is consistent with the region-resolved light curves in Sect.~\ref{sec:results:ribbon_lcs}: the southern ribbon is dominated by a small number of long-lived bright kernels that drift slowly along the path, while the northern ribbon contains multiple kernels that brighten episodically at fixed locations along a longer path. The impulsive flare emission therefore does not correspond to a smooth ribbon front, but to spatially fragmented and temporally intermittent energy deposition.

We assessed how the measured scales depend on the detection threshold and detrending procedure. Varying the prominence threshold from 0.05 to 0.25 increases the median kernel separation from $\sim$1.3~Mm, when fainter substructure is included, to $\sim$2.0--2.3~Mm for the brightest kernels, while the widths remain close to 0.4--0.5~Mm. This suggests a hierarchical ribbon structure, with dominant kernels separated by $\sim$1--2~Mm and accompanied by fainter substructure, while the measured widths remain close to the resolution limit. Repeating the detection on the raw profiles changes the median separations by $<0.16$~Mm and widths by $<0.05$~Mm, but reduces the number of detections by $\sim$10\%, preferentially removing kernels from fainter ribbon regions. The Savitzky--Golay detrending therefore improves detection uniformity along the ribbon without setting the measured spatial scales.

\begin{table}[h]
\caption{Characteristic spatial scales of ribbon-kernel and loop-strand fine
structure. Medians with 25th--75th percentile ranges, over all detections.
Resolution limit $= 0.31$~Mm ($2$ pixels).}
\label{tab:scales}
\centering
\begin{tabular}{lccc}
\hline\hline
Feature & Phase & $\tilde{d}$ (Mm) & $\tilde{w}$ (Mm)  \\
\hline
North ribbon kernels (P1)   & Impulsive & $1.69^{+0.62}_{-0.61}$ & $0.51^{+0.20}_{-0.14}$  \\
South ribbon kernels (P2)   & Impulsive & $1.37^{+0.45}_{-0.46}$ & $0.46^{+0.13}_{-0.12}$  \\
Filament FP kernels (P3)    & Impulsive & $1.38^{+0.61}_{-0.61}$ & $0.42^{+0.29}_{-0.12}$  \\
Post-flare loop strands     & Gradual   & $1.28^{+0.49}_{-0.48}$ & $0.43^{+0.18}_{-0.12}$  \\
\hline
\end{tabular}
\tablefoot{$\tilde{d}$: median peak-to-peak separation along the ribbon (P1--P3)
or between adjacent strands (loops). $\tilde{w}$: median half-prominence width.
Uncertainties give the 25th--75th percentile ranges. Fraction of widths exceeding
the $0.31$~Mm resolution limit: $87\%$ (P1), $81\%$ (P2), $74\%$ (P3), and
$75\%$ (loops); widths below the limit are upper limits. Number of separation /
width measurements: $370/429$ (P1), $60/118$ (P2), $89/103$ (P3), $1514/1954$
(loops).}
\end{table}

\begin{figure}
    \centering
    \includegraphics[width=0.9\columnwidth]{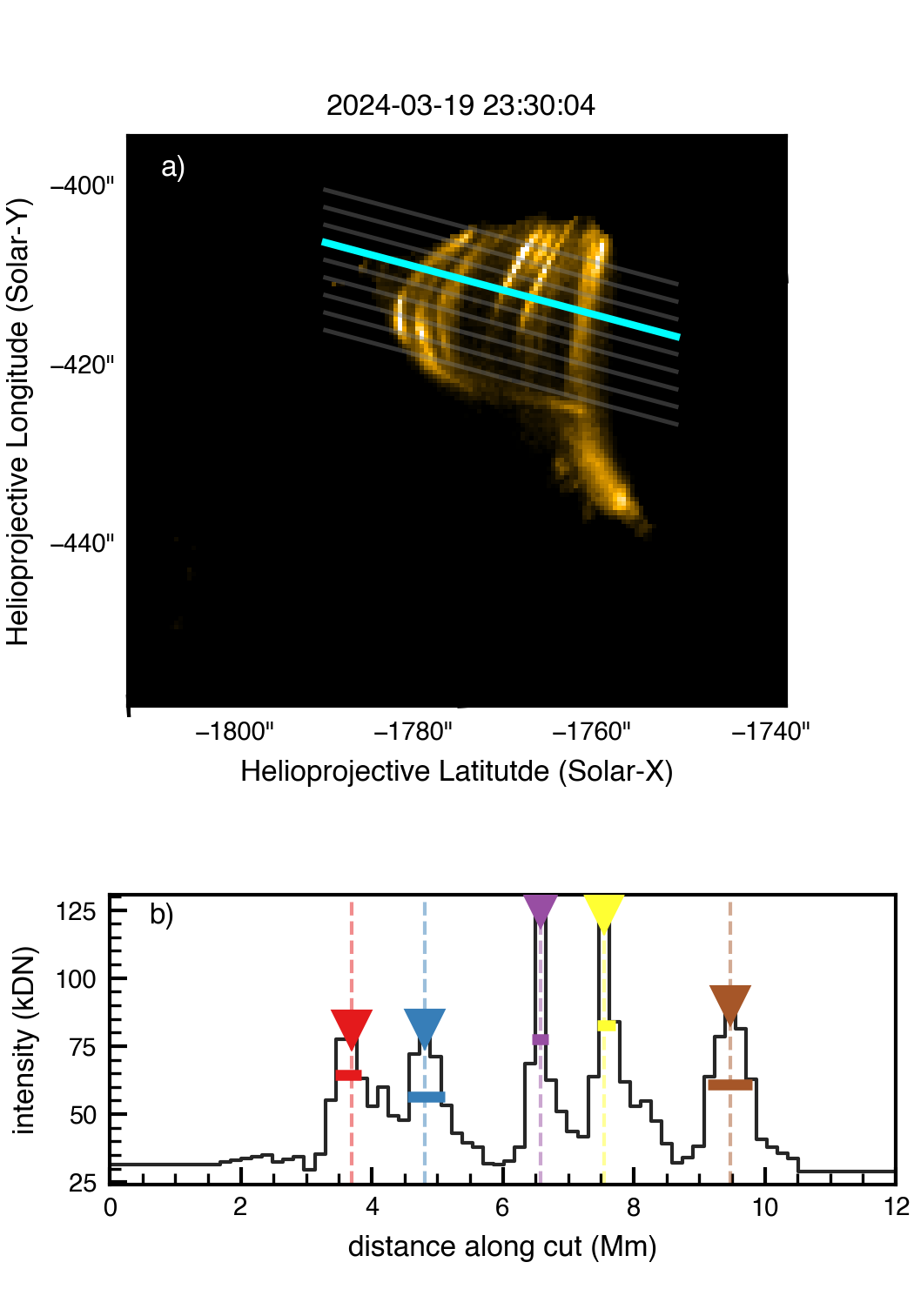}
    \caption{Gradual-phase flare loop arcade. (a) Solar Orbiter/EUI HRI$_{\rm EUV}$ 174~\AA\ image of the arcade during the decay phase. Grey lines show the nine parallel cross-cuts (inclined at $-15\degr$) used to sample the loops transversely; the cyan line marks the cross-cut shown in panel~(b). (b) Intensity profile along the featured cross-cut. Individual strands, detected as peaks of fractional prominence $\geq0.08$, are marked by coloured triangles and dashed vertical lines, with horizontal bars indicating their half-prominence width.}
    \label{fig:slice_decay}
\end{figure}

The same short-exposure observations also resolve the fine structure of the developing post-flare arcade\label{sec:results:loops}. As magnetic reconnection continues into the gradual phase (23:22:48--23:34:48~UT), post-reconnection loop strands accumulate in the growing arcade and brighten in the \hrieuv\ 174~\AA\ passband as the plasma cools through its temperature-response range, allowing us to examine the fine structure of the post-impulsive flare arcade loops, and to compare the loop-strand scales with the impulsive ribbon-kernel measurements from Sect.~\ref{sec:results:kernels}. To measure the transverse strand spacing and width, we placed nine parallel cross-cuts approximately perpendicular to the loop legs (Fig.~\ref{fig:slice_decay}) and identified strands as local intensity peaks using the same prominence-based detection and half-prominence width definition as for the kernels. Here, no detrending is required, since a cut perpendicular to the strands has an approximately flat background.

The post-impulsive flare loops are resolved into multiple narrow strands rather than a smooth loop bundle (Fig.~\ref{fig:slice_decay}). Across all nine cuts and gradual-phase frames, the strands show a characteristic separation of ${\sim}1.3$~Mm and width of ${\sim}0.4$~Mm (Table~\ref{tab:scales}; full distributions in Fig.~\ref{fig:slice_decay_stats}, Appendix~\ref{app:distributions}).  These scales are consistent across height, indicating that the fine structure is not dominated by geometric expansion of the loop bundle. The loop-strand separation shows a similar threshold dependence, increasing from $1.12$ to $1.61$~Mm as the prominence threshold is raised from $0.05$ to $0.25$, whereas the width remains nearly unchanged at $0.42$--$0.51$~Mm. We therefore compare the kernel and loop-strand measurements below using the common threshold of $0.08$.

As for the kernels, the loop-strand widths lie close to the resolution limit: ${\sim}75\%$ exceed the two-pixel limit of 0.31~Mm, but the median corresponds to only about three pixels, so the strands may not be fully resolved. The strand separations, however, are several times the resolution limit.

\begin{figure}
    \centering
    \includegraphics[width=0.8\columnwidth]{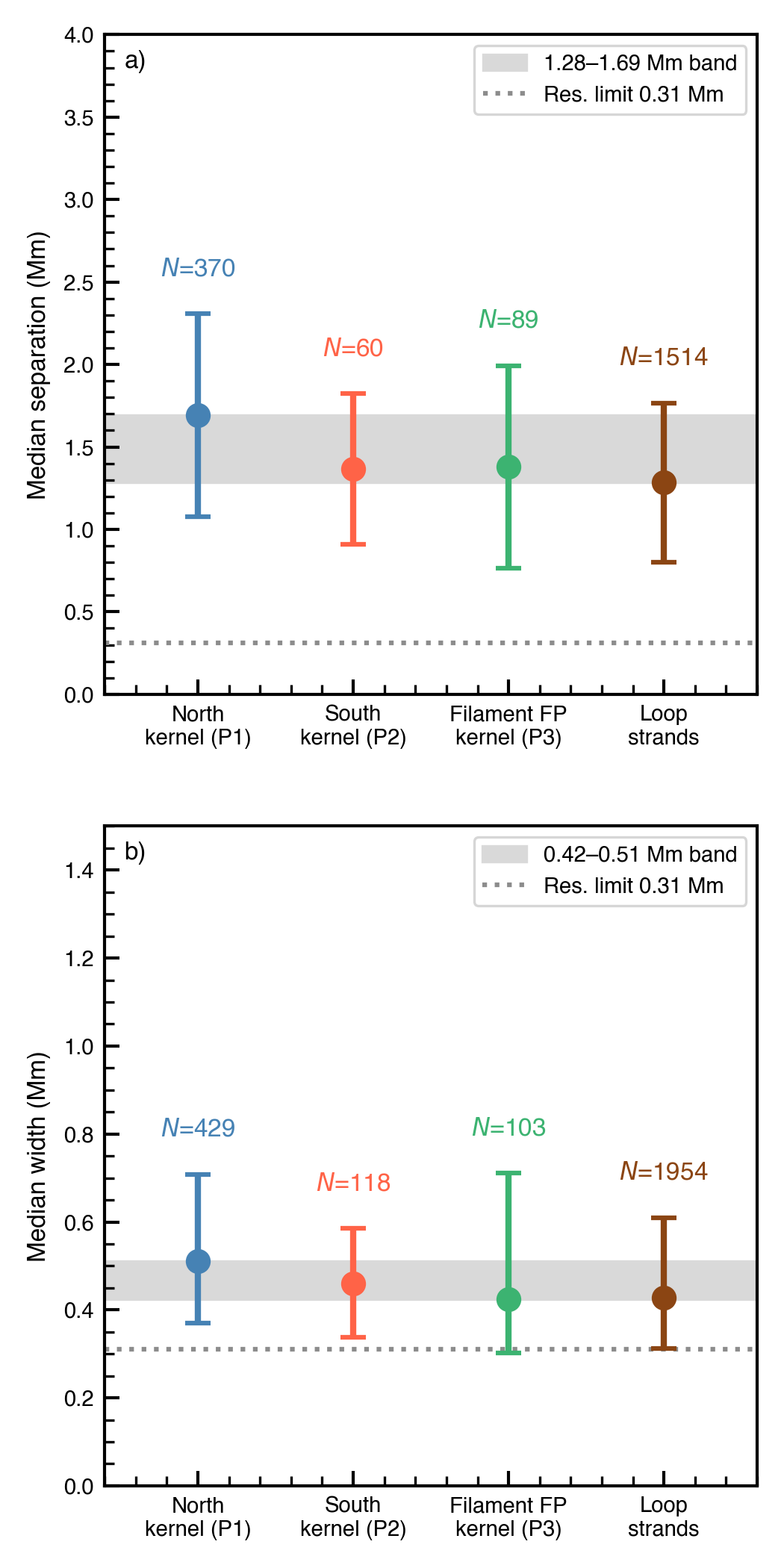}
    \caption{Characteristic spatial scales of the fine structure. Median
    separation (a) and width (b) of the impulsive-phase ribbon kernels (P1, P2, P3)
    and the gradual-phase loop strands; error bars span the interquartile range
    and $N$ is the number of detections. The grey band marks the min--max range
    of the four medians, and the dotted line the two-pixel resolution limit
    (0.31~Mm).}
    \label{fig:bar_plot_sizes}
\end{figure}

Strikingly, the flare loop-strand separation closely matches the impulsive ribbon-kernel separations. This comparison is summarised in Fig.~\ref{fig:bar_plot_sizes}, which collects the median separations and widths of the three ribbon paths and the loop strands into a single view. The median loop-strand separation of 1.28~Mm lies just below the ribbon-kernel medians of 1.37--1.69~Mm, and all four values cluster within a narrow band of approximately 1.3--1.7~Mm. At $4$--$5$ times the two-pixel resolution limit, these separations are the most robust spatial scale measured in this work. The widths show a similar correspondence: the median loop-strand width of 0.43~Mm is comparable to the ribbon-kernel widths of 0.42--0.51~Mm. However, as these characteristic widths lie close to the resolution limit, their intrinsic widths may be smaller than those measured here.

Figure~\ref{fig:bar_plot_sizes} therefore provides the central summary of the fine-structure analysis: the impulsive energy-deposition pattern and the later flare arcade share a common characteristic spatial scale. This agreement is notable because the measurements are made at different flare phases, in different magnetic structures, and using independent profile geometries, suggesting that the correspondence is not simply a consequence of the measurement procedure.

This persistence of comparable spatial scales from the impulsive phase into the formation of the arcade suggests that the flare remains spatially organised throughout. One possible interpretation is fragmented or spatially intermittent reconnection: three-dimensional magnetic models show that fragmented reconnection naturally produces fine structure within ribbons \citep{wyper_2021}, and high-resolution observations reveal complex, rapidly evolving ribbon substructure interpreted in terms of intermittent reconnection and current-sheet dynamics \citep[e.g.][]{nishizuka_2009, french_2021}. In particular, \citet{french_2025} measured a comparable bead spacing of 1.7--1.9~Mm in a different \hrieuv\ Major Flare SOOP observation whose evolution was consistent with tearing-mode behaviour. The comparable scale here, together with the repeated reactivation of preferred kernel locations, is consistent with energy release occurring in spatially localised, recurrent episodes. However, the mapping between current-sheet structure, ribbon emission, and the visible loop arcade is not unique, so the measured separation should be regarded as an empirical characteristic scale of the flare structure rather than a direct measurement of a tearing-mode wavelength. In this context, \cite{sun_2026} recently reported spatially separated, downward-propagating fine-scale structures above flare ribbons in \hrieuv 174~\AA\ observations. Although these downflows differ morphologically from the recurrent along-ribbon kernels studied here, both phenomena reveal rapidly evolving and spatially fragmented ribbon-associated emission. They may therefore share an origin in intermittent energy and momentum deposition associated with fragmented reconnection, although the present observations do not allow us to establish a direct physical connection. Such intermittency is also relevant to models of quasi-periodic pulsations \citep{mclaughlin_2018}, but no periodicity analysis is performed here and the observed variability should not be interpreted as a QPP detection.

The widths should be interpreted more cautiously than the separations. Their medians of 0.4--0.5~Mm lie close to the two-pixel \hrieuv\ resolution limit, so the intrinsic widths may extend to smaller scales. They are broadly consistent with coronal-strand widths measured with the High-resolution Coronal Imager (Hi-C) although the narrowest Hi-C strands remain smaller \citep{williams_2020}. Ground-based observations show considerably finer structure below the \hrieuv\ resolution: NST reports ribbon, loop, and footpoint widths of ${\approx}80$--200~km \citep{sharykin_2014, jing_2016}, while DKIST resolves chromospheric ribbon blobs of a few hundred kilometres \citep{yadav_2025} and post-flare loop strands of only a few tens of kilometres \citep{tamburri_2025}, so at least some \hrieuv\ kernels and strands likely contain unresolved substructure. Line-of-sight superposition and the near-limb geometry add further uncertainty, since an apparently discrete strand may be the projection of a more complex three-dimensional structure \citep{malanushenko_2022}. The conclusion is therefore not the intrinsic width of the kernels or strands, but the persistence of fine spatial organisation from the impulsive ribbons into the early arcade, which constrains how localised reconnection and energy deposition build the larger-scale flare-loop system.

\subsection{Footpoint area and the implied non-thermal energy flux}
\label{sec:results:energy}
The close temporal correspondence between the \hrieuv\ short-exposure kernels and the STIX HXR emission motivates using the compact EUV kernel area as a proxy for the local energy-deposition scale. We emphasise, however, that this is an observational proxy as the 174~\AA\ emission traces the atmospheric response rather than the electron beam directly, and its morphology depends on the passband temperature response, plasma density, and possible conductive or Alfv\'enic heating. The EUV-emitting area therefore need not be identical to the area over which non-thermal electrons deposit their energy. Accordingly, the energy fluxes derived using the \hrieuv\ area should be interpreted as conditional estimates, under the assumption that the beam-impact area is comparable to the compact EUV-emitting area. Nevertheless, the high-resolution \hrieuv provides a complementary constraint to HXR and UV imaging, whose apparent source areas need not correspond directly to the beam-impact area. HXR source sizes are constrained by the spatial resolution and indirect Fourier-imaging reconstruction, and can also be affected by electron transport through the atmosphere and X-ray albedo \citep{dennis_2009,kontar_2010,battaglia_2012,massa_2023}, while AIA~1600~\AA\ ribbon emission can extend beyond the directly heated core \citep{simoes_2019}. We therefore compare the \hrieuv, AIA~1600~\AA, and STIX source areas at the two main HXR peaks, centred on 23:19:57 (Peak~1) and 23:20:22~UT (Peak~2; shaded in Fig.~\ref{fig:lightcurves}), and assess how the choice of area affects the inferred non-thermal energy flux.


The non-thermal electron power, $P_{\rm nth}$, at each peak is derived from STIX spectral fitting. The spatially integrated spectra were fitted with OSPEX over ${\approx}5$--60~keV using an isothermal (\texttt{vth}) plus thick-target (\texttt{thick2}) model with an albedo correction \citep{brown_1971, holman_2011}; the fitted spectra and residuals are shown in Appendix~\ref{app:stix_spectra} (Fig.~\ref{fig:fitted_spectra}). 

Peak~1 is described by a 16.7~MK thermal component and a non-thermal spectrum with power-law index $\delta=5.86$ and low-energy cut-off $E_c=19.8$~keV, giving $P_{\rm nth}=(3.4\pm0.8)\times10^{27}$~erg~s$^{-1}$. Peak~2 has a harder non-thermal spectrum, with $\delta=4.33$ and $E_c=17.6$~keV, and requires an additional super-hot thermal component at 32.0~MK, giving a best-fit non-thermal power of $P_{\rm nth}=1.4\times10^{27}$~erg~s$^{-1}$. At this peak, $E_{\rm c}$ is not well constrained and fits across the tested range $E_{\rm c}=10$--$30$~keV are statistically comparable and give $P_{\rm nth}=(0.4$--$5.0)\times10^{27}$~erg~s$^{-1}$.

\begin{figure*}
    \centering
    \includegraphics[width=0.8\textwidth]{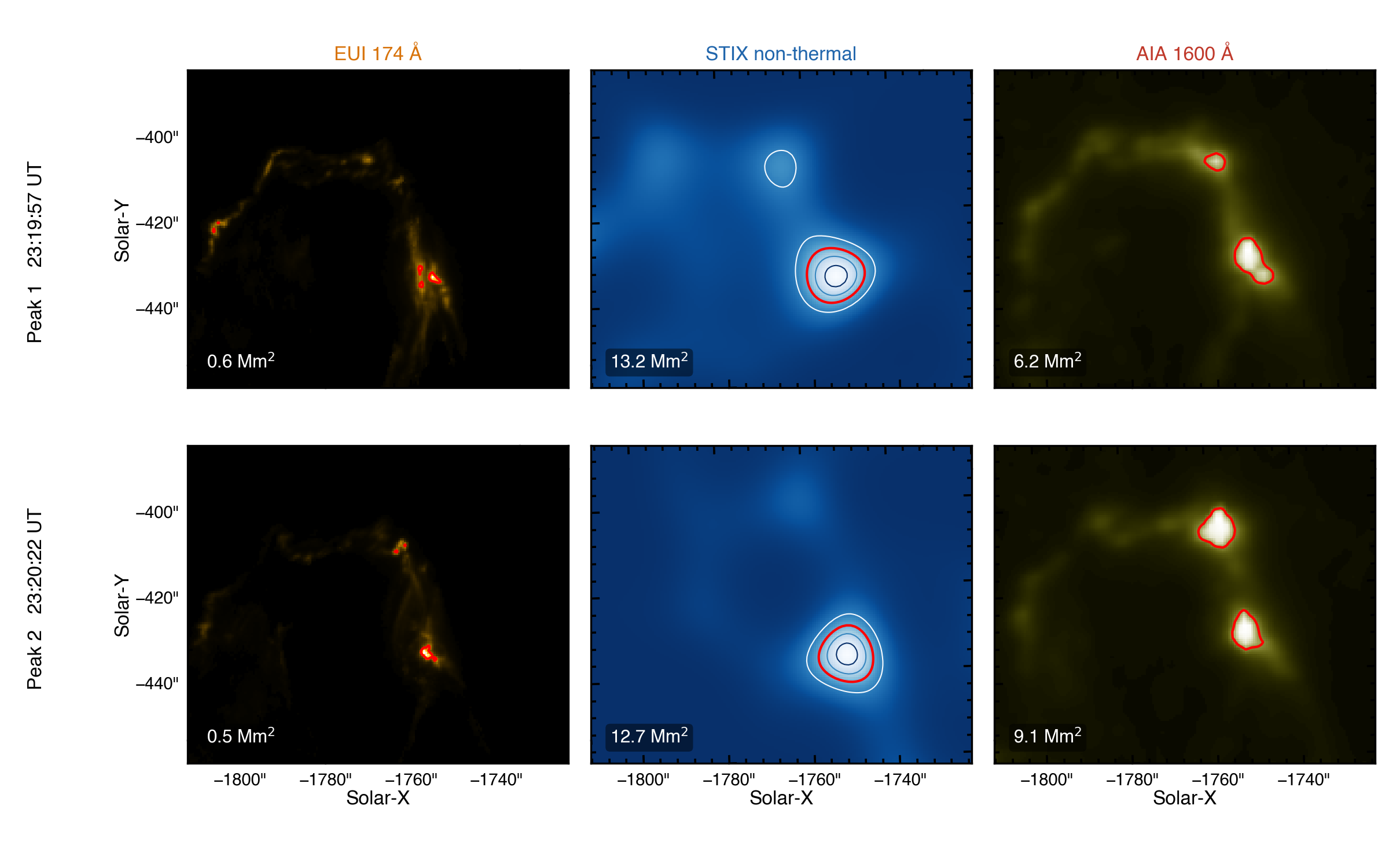}
    \caption{Footpoint size comparison at the two hard X-ray peaks.
    Rows show Peak~1 (23:19:57~UT) and Peak~2 (23:20:22~UT). Columns
    show the \hrieuv\ 174~\AA\ short-exposure emission, the STIX
    non-thermal hard X-ray source, and the AIA~1600~\AA\ emission.
    AIA and STIX images are shown reprojected onto the Solar Orbiter
    viewpoint for visual comparison. Red contours mark the
    50\%-of-peak level used to define the footpoint area, and the
    corresponding area is annotated in each panel. The \hrieuv\
    emission isolates substantially more compact footpoint structure
    than either AIA~1600~\AA\ or STIX.}
    \label{fig:footpoint_maps}
\end{figure*}

For each instrument, we define the footpoint area as the area enclosed by the 50\%-of-peak intensity contour,
\[
A_{50} = N_{\rm pix}(I > 0.5 I_{\rm max})~A_{\rm pix}.
\]
where $N_{\rm pix}$ is the number of pixels with intensity $I>0.5I_{\rm max}$, $I_{\rm max}$ is the peak intensity, and $A_{\rm pix}$ is the physical area per pixel. For \hrieuv, the area is measured from short-exposure images summed over the corresponding STIX spectral fitting interval. For AIA~1600~\AA, we use the nearest frame after correcting for the light travel travel time difference between Solar Orbiter and SDO. For STIX, we use the reconstructed non-thermal HXR source. Areas are computed from the native data using the appropriate physical pixel scale, with the AIA images additionally reprojected onto the Solar Orbiter viewpoint for the visual comparison in Fig.~\ref{fig:footpoint_maps}.

Figure~\ref{fig:footpoint_maps} shows that the \hrieuv\ short exposures isolate far more compact footpoint emission than either AIA~1600~\AA\ or STIX. At the 50\% level the \hrieuv\ areas are 0.65 and 0.53~Mm$^2$ at Peaks~1 and 2, compared with 6.2 and 9.1~Mm$^2$ for AIA~1600~\AA\ and 13.2 and 12.7~Mm$^2$ for STIX (Table~\ref{tab:footpoint}).  The \hrieuv\ footpoint is approximately an order of magnitude smaller than the AIA ribbon and ${\sim}20$ times smaller than the STIX source, isolating a single compact kernel embedded within the broader AIA ribbon, consistent with the hierarchy above and supporting the use of the compact kernel area as a practical proxy for the local deposition scale.  This ordering holds across peak intensity thresholds of 30--70\% (Appendix~\ref{app:areathreshold}). The STIX areas are limited by the effective angular resolution of the image reconstruction and should be treated as upper limits on the intrinsic HXR footpoint scale. 

\begin{table}
\caption{Footpoint areas at the 50\% intensity level and implied
non-thermal energy fluxes at the two hard X-ray peaks.}
\label{tab:footpoint}
\centering
\begin{tabular}{lcccc}
\hline\hline
 & \multicolumn{2}{c}{$A_{50}$ (Mm$^2$)}
 & \multicolumn{2}{c}{$F_{\rm nth}$ (erg~cm$^{-2}$~s$^{-1}$)} \\
Instrument & Peak 1 & Peak 2 & Peak 1 & Peak 2 \\
\hline
\hrieuv\ 174~\AA     & 0.65 & 0.53 & $5.2\times10^{11}$ & $2.6\times10^{11}$ \\
AIA 1600~\AA         & 6.2  & 9.1  & $5.5\times10^{10}$ & $1.5\times10^{10}$ \\
STIX non-thermal      & 13.2 & 12.7 & $2.6\times10^{10}$ & $1.1\times10^{10}$ \\
\hline
\end{tabular}
\tablefoot{$F_{\rm nth}=P_{\rm nth}/A_{50}$, using best-fit powers
$P_{\rm nth}=3.4\times10^{27}$ and
$1.4\times10^{27}$~erg~s$^{-1}$ for Peaks~1 and 2, respectively.
The tabulated fluxes are nominal values at the 50\% area threshold. The intensity-threshold sensitivity is shown in in Fig.~\ref{fig:energy_flux}, while the while the spectral-model uncertainty is discussed in Sect.~\ref{sec:results:energy}.}
\end{table}

\begin{figure*}
    \centering
    \includegraphics[width=0.9\textwidth]{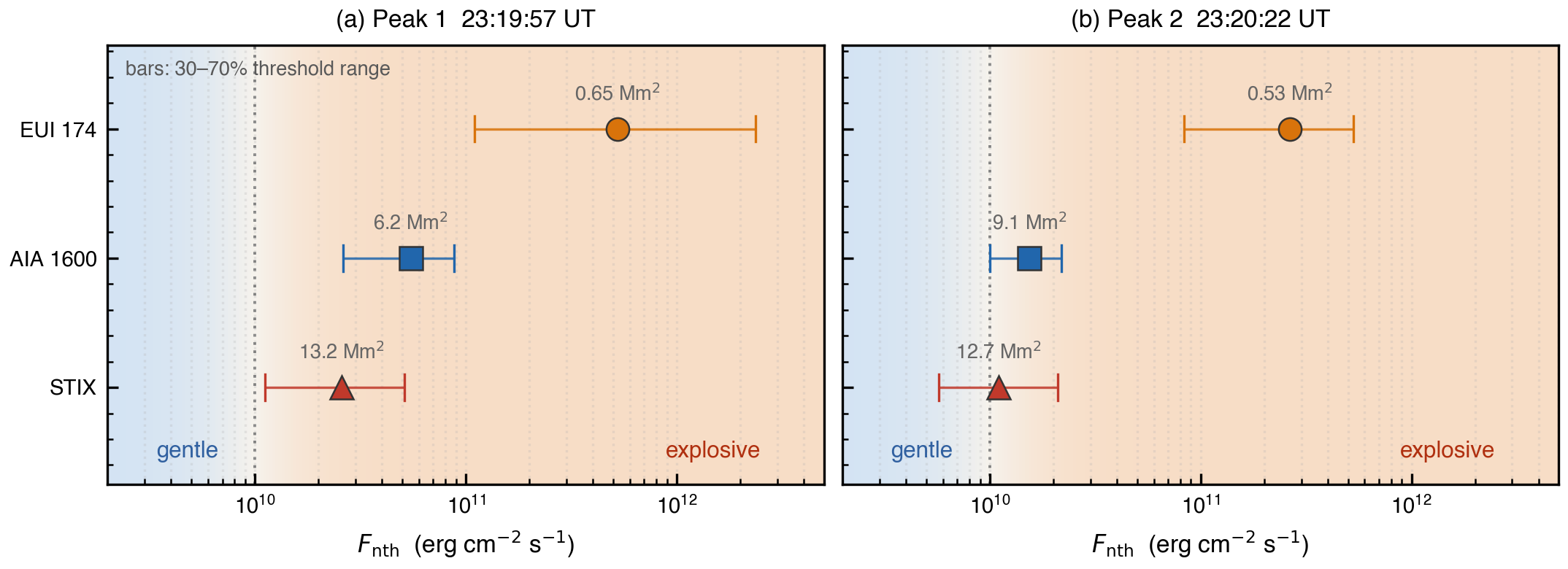}
    \caption{Implied non-thermal energy flux,
    $F_{\rm nth}=P_{\rm nth}/A_{50}$, for \hrieuv, AIA~1600~\AA, and
    STIX footpoint areas at (a) Peak~1 and (b) Peak~2. The background
    indicates the nominal ${\sim}10^{10}$~erg~cm$^{-2}$~s$^{-1}$
    scale often associated with the transition from gentle to explosive
    evaporation \citep{fisher_1985a, fisher_1985b}; it is shown as a
    gradual transition because this is an order-of-magnitude criterion
    rather than a sharp boundary. Bars span the fluxes obtained over the 30–70\% intensity-threshold range of the measured footpoint area (Appendix~\ref{app:areathreshold}) using the best-fit electron powers with markers show the 50\% reference value. The compact \hrieuv\ footpoint areas imply substantially larger local energy fluxes than the AIA or STIX areas across the full threshold range.}
    
    \label{fig:energy_flux}
\end{figure*}

The consequence for the inferred non-thermal energy flux, $F_{\rm nth}$, is substantial. Using $F_{\rm nth}=P_{\rm nth}/A_{50}$, Fig.~\ref{fig:energy_flux} translates these area measurements and their threshold dependence into inferred local energy fluxes. The markers show the nominal 50\% values, while the bars span the values obtained over thresholds of 30--70\%. For the best-fit electron powers, the \hrieuv-based estimates are of order $10^{11}$--$10^{12}$~erg~cm$^{-2}$~s$^{-1}$, compared with values of order $10^{10}$~erg~cm$^{-2}$~s$^{-1}$ from AIA and STIX, and remain substantially higher at every matched threshold. The broad \hrieuv\ range reflects the sharply peaked kernel emission. Uncertainty in $P_{\rm nth}$ broadens the absolute ranges but shifts all three instrument-based estimates together. Thus, for the same electron power, using the compact footpoint scale substantially increases the inferred local energy flux.

This difference changes the physical interpretation. Because source areas inferred from AIA ribbons or HXR sources can substantially exceed the spatial scale of the actual footpoint area, flare energy fluxes estimated from UV, H$\alpha$, white-light, or HXR ribbon areas \citep[e.g.][]{krucker_2011} can underestimate the local flux for compact impulsive kernels. This was anticipated from RHESSI and TRACE studies of compact footpoints \citep{fletcher_2007, krucker_2011} and sub-arcsecond IRIS kernels \citep{graham_2015}. For example, \citet{krucker_2011} derived a local deposition rate exceeding $10^{12}$~erg~cm$^{-2}$~s$^{-1}$ using a resolved optical ribbon width. The contribution here is to measure the compact the EUV response area directly during the HXR peak, without saturation, providing a higher-resolution constraint on the spatial scale relevant to the inferred local beam energy flux.

The choice of footpoint area also affects where the inferred energy flux lies relative to commonly used evaporation thresholds, as illustrated in Fig.~\ref{fig:energy_flux}. With AIA- or STIX-based areas the inferred energy fluxes at both peaks fall near the ${\sim}10^{10}$~erg~cm$^{-2}$~s$^{-1}$ scale conventionally associated with the transition from gentle to explosive chromospheric evaporation \citep{fisher_1985a, fisher_1985b}, whereas the \hrieuv\-based estimates lie approximately an order of magnitude above it. This threshold should be regarded only as an order-of-magnitude guide: it derives from one-dimensional loop models with specific beam parameters, and the transition depends on the low-energy cutoff, spectral index, heating duration, and pre-flare atmosphere \citep{reep_2015}, while observations show both regimes, and transitions between them, across and within events \citep{milligan_2006}. We therefore do not interpret the \hrieuv-based fluxes as direct evidence of explosive evaporation, which would require co-temporal footpoint spectroscopy of blueshifted high-temperature lines \citep[e.g.][]{polito_2016}, but rather, measuring the compact footpoint area shifts the inferred local energy flux from near the conventional transition scale to the range commonly associated with explosive evaporation.

This has implications for radiative-hydrodynamic flare modelling, in which the beam energy flux is a principal input parameter \citep{allred_2015}. Commonly adopted model grids span ${\sim}10^{9}$--$10^{11}$~erg~cm$^{-2}$~s$^{-1}$, with fluxes up to $10^{13}$~erg~cm$^{-2}$~s$^{-1}$ invoked to reproduce continuum observations \citep{kowalski_2017}. Here, however, we found fluxes exceeding $10^{11}$~erg~cm$^{-2}$~s$^{-1}$ onto sub-Mm$^2$ areas are inferred even for a moderate M-class flare once the compact footpoint scale is resolved. Because the kernels may themselves contain unresolved substructure (Sect.~\ref{sec:results:kernels}), the \hrieuv-based value may still underestimate the peak local flux, further motivating multithread treatments in which observed footpoints comprise sequentially heated sub-resolution threads \citep{reep_2020}. More generally, the spatial fragmentation of the energy deposition means that a single flux need not characterise an entire ribbon: position-dependent heating histories are seen in IRIS observations and modelling of ribbon fronts \citep{polito_2023}, suggesting that model comparisons should be made at kernel scales and on correspondingly short timescales \citep{collier_2026}.

\section{Conclusions}
\label{sec:conclusions}
We have analysed the 2024 March 19 M2.1 flare observed with the dedicated short-exposure \hrieuv\ flare-watch mode during the Solar Orbiter Major Flare SOOP. The 0.04~s exposures, obtained at an effective cadence of 2~s, provide access to the compact and rapidly evolving flare emission that is largely obscured by saturation in the normal-exposure \hrieuv\ and AIA images. Combined with STIX timing, imaging, and spectroscopy, these observations allow us to connect the spatial organisation of the impulsive flare emission with both the developing post-flare arcade and the inferred local energy deposition. Our main conclusions are:

\begin{itemize}

\item The spatially integrated short-exposure \hrieuv\ emission closely follows the non-thermal STIX HXR emission during the impulsive phase. The rapid EUV and HXR variability is co-temporal within the 2~s \hrieuv\ sampling, supporting the interpretation of the compact 174~\AA\ kernels as tracers of the impulsive lower-atmosphere response. At the same time, the region-resolved \hrieuv\ observations show that this response is highly structured, with different ribbon regions exhibiting distinct temporal evolution.

\item The impulsive ribbons are composed of compact, repeatedly activated kernels with characteristic separations of ${\sim}1.4$--$1.7$~Mm. The developing post-flare arcade shows a remarkably similar strand separation of ${\sim}1.3$~Mm. In contrast, the measured kernel and strand widths of ${\sim}0.4$--$0.5$~Mm lie close to the \hrieuv\ resolution limit and may extend to smaller intrinsic scales. The persistence of the ${\sim}1$--$2$~Mm spatial organisation from the impulsive ribbons into the newly formed arcade provides an observational constraint on how spatially localised energy release maps into the subsequent flare-loop system.

\item Combining the compact \hrieuv\ footpoint areas (${\sim}0.5$--$0.65$~Mm$^2$) with STIX spectroscopy yields inferred local non-thermal energy fluxes approximately an order of magnitude higher than estimates based on AIA or STIX source areas, with nominal values of a few $\times10^{11}$~erg~cm$^{-2}$~s$^{-1}$ at the two peaks. These higher kernel-scale fluxes should therefore be accounted for in radiative-hydrodynamic flare modelling, where unresolved or ribbon-averaged footpoint areas may substantially underestimate the local energy flux.

\end{itemize}

The Major Flare SOOP has now captured additional flares in the short-exposure \hrieuv\ mode \citep[see][]{ryan_2025}, providing an opportunity to test whether the compact kernel scales, ${\sim}1$--$2$~Mm spatial organisation, and enhanced local energy fluxes found here are common properties of flares or depend on flare magnitude, magnetic topology, or evolutionary phase. Coordinated spectroscopy from SPICE and IRIS, and future facilities such as the Multi-slit Solar Explorer (MUSE) \citep{de_pontieu_2020} and Solar-C/EUVST \citep{shimizu_2019}, will be essential for linking this compact morphology to plasma dynamics, chromospheric evaporation, and energy transport at comparable spatial and temporal scales.

More broadly, these observations provide a compelling proof of concept for flare-focused EUV imaging. Alternating short and normal exposures extends the usable dynamic range while retaining the spatial resolution and rapid cadence needed to capture compact flare structure and its rapid evolution together with the surrounding coronal context. In this event, that combination reveals spatial scales and local energy fluxes that would otherwise be obscured or underestimated. These results highlight the value of flare-optimised EUV imaging that combines high dynamic range, high spatial resolution, and high cadence, particularly when paired with co-temporal spectroscopy, and motivate future instrumentation such as the High Resolution Flare Imager (HiFI) concept for the Solar Particle Acceleration, Radiation \& Kinetics (SPARK) mission proposed to the ESA M8 call.

\begin{acknowledgements}
Solar Orbiter is a space mission of international collaboration between ESA and NASA, operated by ESA. The STIX instrument
is an international collaboration between Switzerland, Poland, France, Czech Republic, Germany, Austria, Ireland, and Italy. L.A.H is supported by a Royal Society-Research Ireland University Research Fellowship (URF$\backslash$R1$\backslash$241775). S.K. and H.C. are supported by the Swiss National Science Foundation Grant 200021L\_189180 for STIX. The EUI instrument was built by CSL, IAS, MPS, MSSL/UCL, PMOD/WRC, ROB, LCF/IO with funding from the Belgian Federal Science Policy Office (BELSPO/PRODEX PEA 4000112292 and 4000134088); the Centre National d’Etudes Spatiales (CNES); the UK Space Agency (UKSA); the Bundesministerium für Wirtschaft und Energie (BMWi) through the Deutsches Zentrum für Luft- und Raumfahrt (DLR); and the Swiss Space Office (SSO). The authors thank the anonymous referee for their constructive comments and feedback which improved this manuscript.
\end{acknowledgements}

\bibliographystyle{aa} 
\bibliography{ananda.bib} 

\begin{appendix}
\section{Co-registration, combined images, and compression}
\label{app:processing}
The normal- and short-exposure sequences were reprojected onto a common helioprojective grid using the the World Coordinate System (WCS) of the first normal-exposure image as the reference. The reprojection propagates coordinates with the solar surface, removing the apparent motion due to solar rotation and applying the pointing information in the level-2 WCS metadata, which corrects the known frame-to-frame pointing variations. A small residual jitter remains but is below one pixel between frames. Because the short-exposure images contain many undefined pixels outside the transmitted bright regions, direct full-frame cross-correlation is not robust, so no additional empirical jitter correction is applied. Since the kernel and loop measurements are made from individual frames and local intensity profiles, this sub-pixel jitter does not affect the results.

For visualisation only, we construct combined images from the co-registered normal- and short-exposure pairs: pixels in the saturated flare core of the normal-exposure image are replaced by the corresponding short-exposure signal, with both images in the same DN~s$^{-1}$ units. This preserves the faint coronal context from the normal exposure while showing the unsaturated flare core (Figs.~\ref{fig:fsi_hri}~c and \ref{fig:fsi_hri_normal_short}~c). All quantitative measurements of the compact flare emission are made from the short-exposure images themselves. To sustain the observing mode within Solar Orbiter's telemetry constraints, the short-exposure data were transmitted using onboard compression; for the event analysed here the data products were generated using lossless compression.

\section{AIA comparison}
\label{app:aia_comparison}
\begin{figure*}
    \centering
    \includegraphics[width=0.90\textwidth]{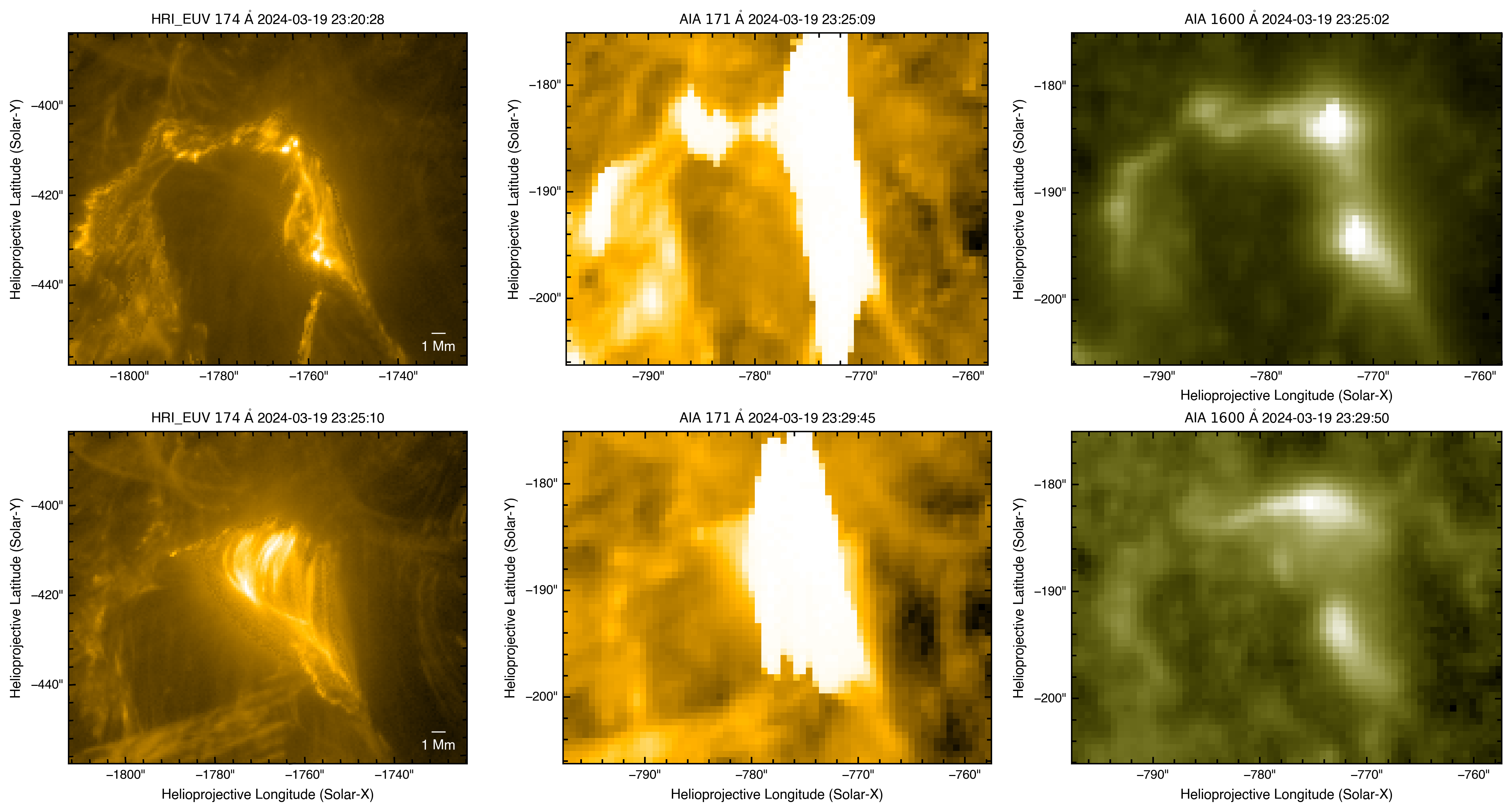}
    \caption{Comparison between EUI/\hrieuv\ and SDO/AIA imaging during the impulsive and early gradual phases. Left column: combined \hrieuv\ images, in which the short-exposure signal replaces the saturated core of the normal-exposure image. Middle column: nearest AIA 171~\AA\ images to the left \hrieuv\ observations. Right column: nearest AIA 1600~\AA\ images to the left \hrieuv\ observations. AIA 171~\AA\ saturates strongly in the flare core, while AIA 1600~\AA\ shows the broader ribbon-scale emission. The combined \hrieuv\ images preserve both the surrounding coronal context and the compact unsaturated flare structure.}
\label{fig:aia_eui_comparison}
\end{figure*}

Figure~\ref{fig:aia_eui_comparison} places the \hrieuv\ observations in the context of the nearest SDO/AIA 171 and 1600~\AA\ images during the impulsive and early gradual phases. The comparison illustrates the strong saturation of the AIA 171~\AA\ flare core and the broader ribbon morphology seen at 1600~\AA, relative to the compact structure visible in the \hrieuv\ short-exposure images.

\section{Kernel and loop-strand spatial scales.}
\label{app:distributions}
\begin{figure*}
    \centering
    \includegraphics[width=0.8\textwidth]{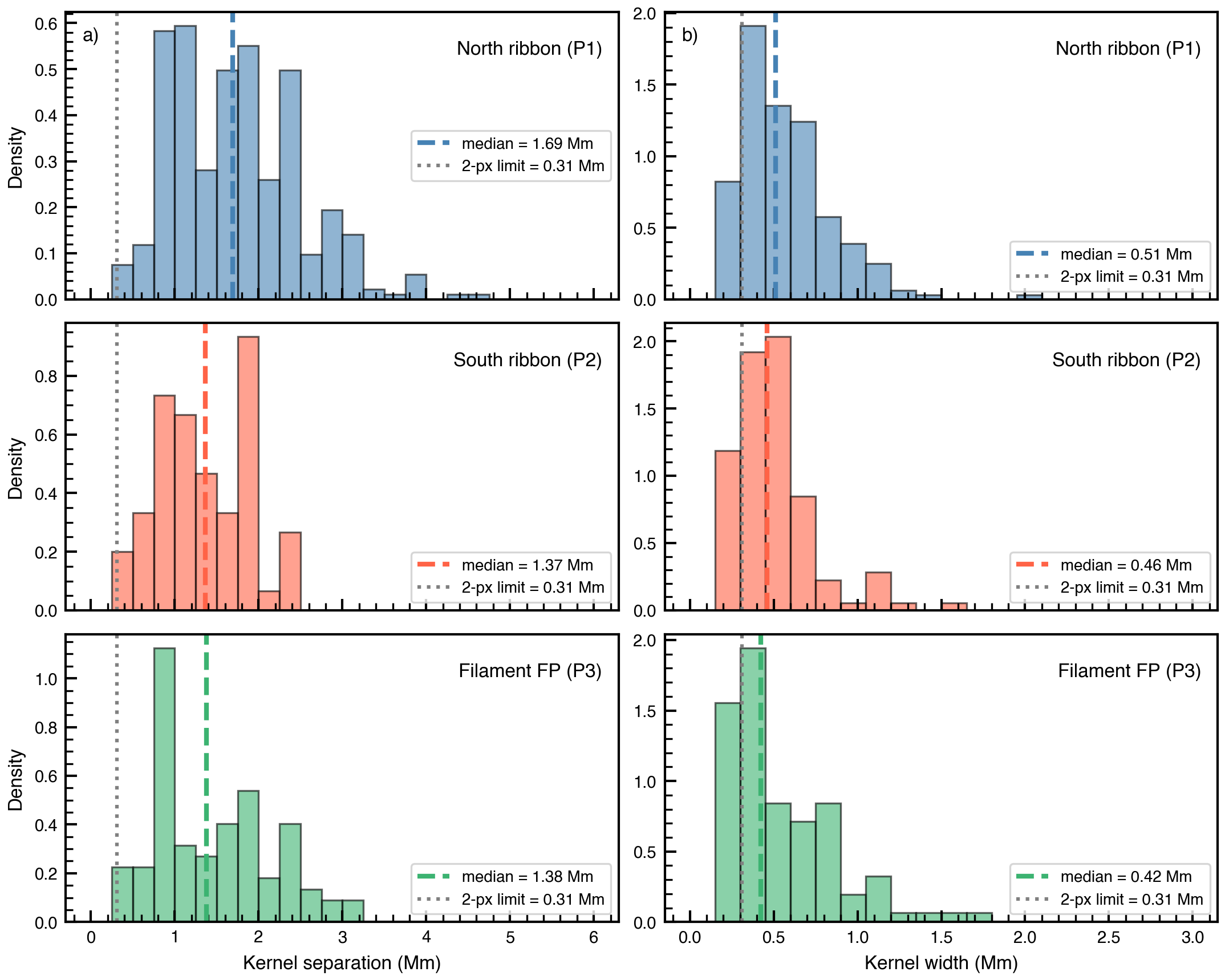}
    \caption{Spatial scales of the impulsive-phase ribbon kernels. Distributions of the (a) separation and (b) half-prominence widths of the bright kernels detected along the North ribbon (P1, blue), South ribbon (P2, red), and filament footpoint (P3, green), pooled over all impulsive-phase frames and normalised to unit area. Dashed vertical lines mark the median of each distribution and the dotted line the two-pixel resolution limit ($0.31$~Mm). The median kernel separations are $1.69$, $1.37$, and $1.38$~Mm, and the median full width at half prominence are $0.51$, $0.46$, and $0.42$~Mm, for P1, P2, and P3, respectively.}
    \label{fig:stats_imp_kernels}
\end{figure*}

\begin{figure*}
    \centering
    \includegraphics[width=0.7\textwidth]{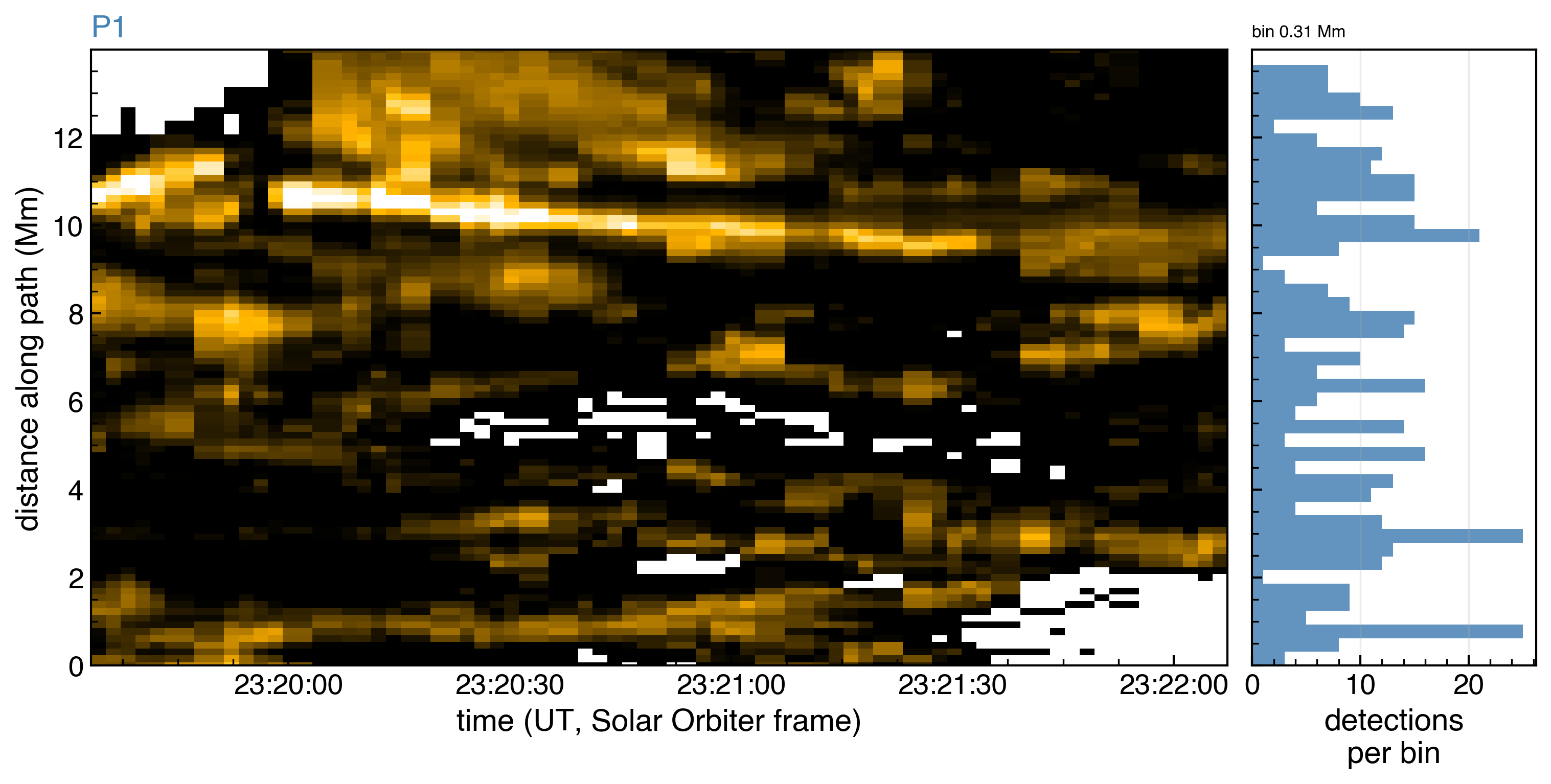}
    \caption{Recurrent kernel brightenings along the northern ribbon (P1). (Left) Space-time map of slit-averaged \hrieuv\ 174~\AA\ intensity along the P1 path, with each column representing one $2$~s short-exposure frame and distance measured along the path in Fig.~\ref{fig:imp_slice}~a. Near-horizontal tracks indicate repeated brightenings at similar positions, the brightest, near $10$~Mm, persists for almost two minutes while drifting slowly along the path. (Right) Distribution of the same detections along the path, binned at the two-pixel resolution limit ($0.31$~Mm) and sharing the vertical axis. }
    \label{fig:spacetime}
\end{figure*}

\begin{figure*}
    \centering
    \includegraphics[width=0.8\textwidth]{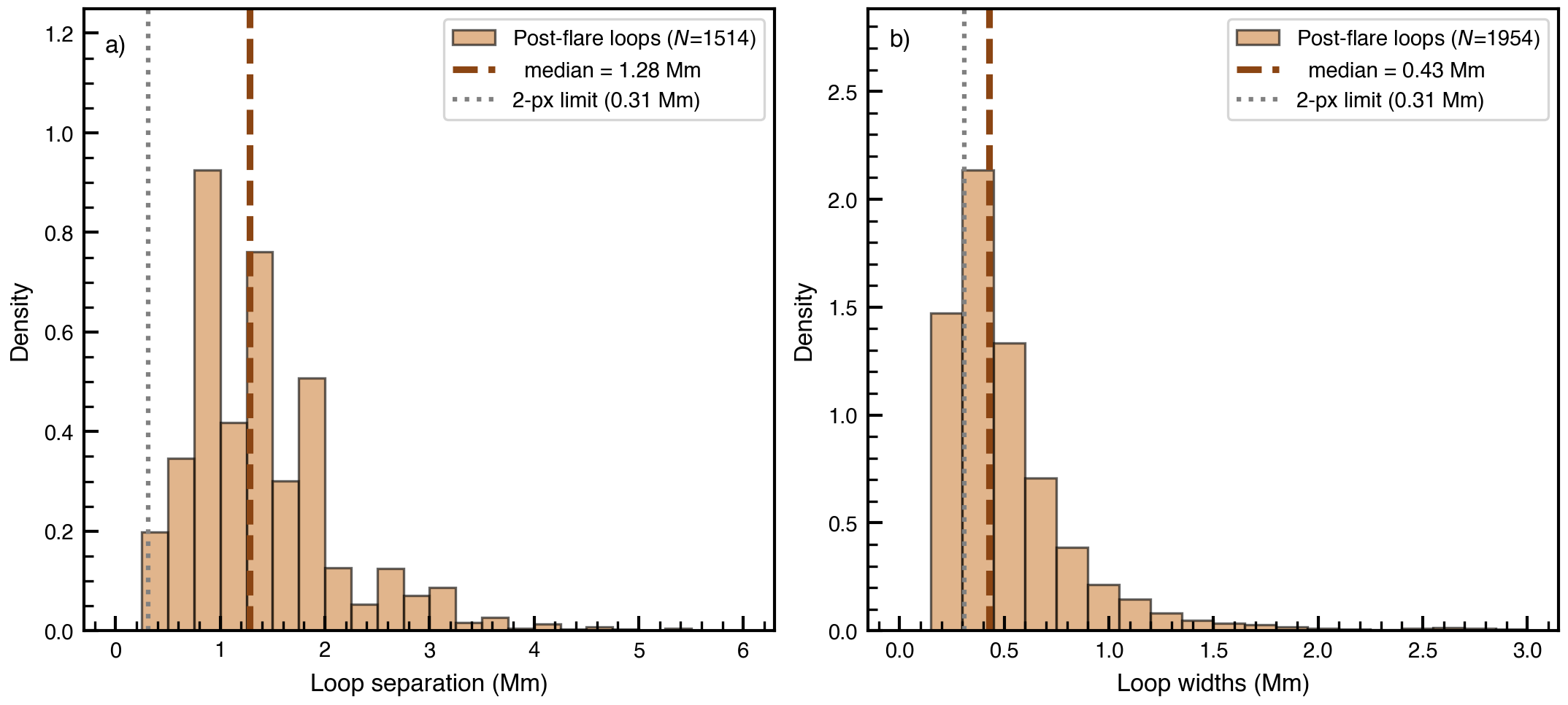}
    \caption{Spatial scales of the gradual-phase loop strands. Distributions of the (a) transverse separation and (b) width of the loop strands detected across all nine cross-cuts and all gradual-phase frames, normalised to unit area. The dashed line marks the median (separation $1.28$~Mm; width $0.43$~Mm) and the dotted line the $2$-pixel resolution limit ($0.31$~Mm).}
    \label{fig:slice_decay_stats}
\end{figure*}

The median separations and widths of the impulsive-phase ribbon kernels and the gradual-phase loop strands quoted in Sect.~\ref{sec:results:kernels}, summarised in Table~\ref{tab:scales} and Fig.~\ref{fig:bar_plot_sizes}, are drawn from the full distributions shown here: Fig.~\ref{fig:stats_imp_kernels} for the ribbon kernels along the three paths (P1--P3) and Fig.~\ref{fig:slice_decay_stats} for the loop strands across the nine cross-cuts.
Figure~~\ref{fig:spacetime} complements these distributions by showing the temporal evolution of the kernel locations along P1, where repeated brightenings appear at similar positions over tens of seconds.

\section{X-ray spectral fitting}
\label{app:stix_spectra}
\begin{figure*}
    \centering
    \includegraphics[width=0.8\textwidth]{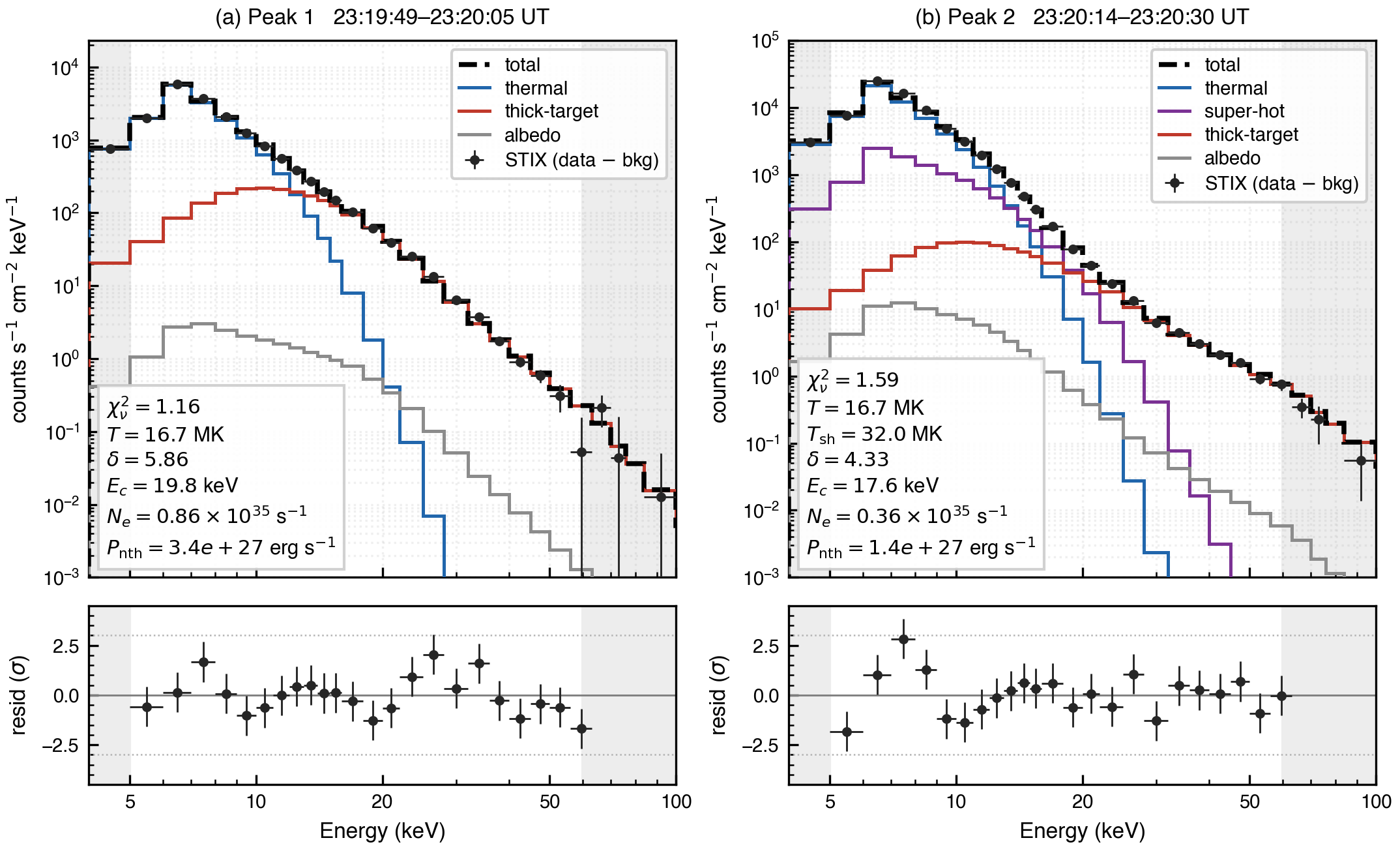}
    \caption{STIX X-ray count spectra at the two hard X-ray peaks:
    (a) Peak~1, fitted over 23:19:49--23:20:05~UT, and
    (b) Peak~2, fitted over 23:20:14--23:20:30~UT. Each spectrum is
    fitted with an isothermal thermal component (blue), a thick-target
    non-thermal component (red), and an albedo correction (grey).
    Peak~2 additionally includes a super-hot thermal component
    (purple). The total model is shown by the dashed black curve and
    the background-subtracted STIX data by black points with
    $1\sigma$ uncertainties. Grey shaded regions lie outside the fitted
    energy range. Lower panels show the normalised residuals.}
    \label{fig:fitted_spectra}
\end{figure*}

The non-thermal electron powers used in Sect.~\ref{sec:results:energy} are derived from the STIX spectral fits shown in Fig.~\ref{fig:fitted_spectra}; the fitted parameters are given in the main text.

\section{Sensitivity of the footpoint areas to the intensity threshold}
\label{app:areathreshold}

\begin{figure*}
    \centering
    \includegraphics[width=\textwidth]{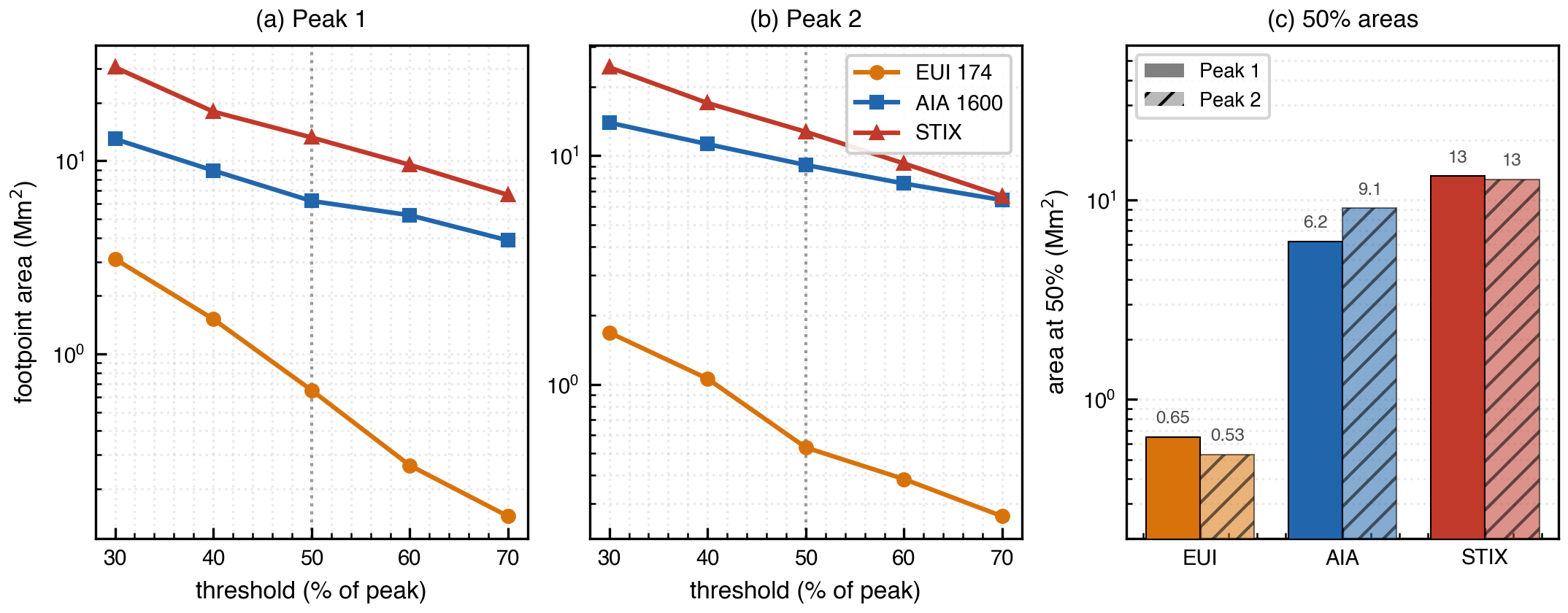}
    \caption{Footpoint area as a function of intensity threshold. (a), (b) Area versus
    threshold ($30$--$70\%$ of the peak intensity) for \hrieuv, AIA~1600~\AA, and STIX,
    at Peak~1 and Peak~2. (c) The 50\%-level areas for all three instruments (Peak~1
    solid, Peak~2 hatched). The \hrieuv\ footpoint is the smallest at every threshold
    and the most threshold-sensitive.}
    \label{fig:area_comparison}
\end{figure*}

The footpoint areas in Sect.~\ref{sec:results:energy} are measured at the 50\% intensity level. Figure~\ref{fig:area_comparison} shows their dependence on the threshold over the range $30$--$70\%$, together with a summary of the 50\% values. The ordering \hrieuv\ $\ll$ AIA $<$ STIX is preserved at every threshold, confirming that the order-of-magnitude compactness of the \hrieuv\ footpoint relative to AIA and STIX is not an artefact of the 50\% choice. The \hrieuv\ area is the most threshold- sensitive, varying by $\approx$20 times across $30$--$70\%$ for the first peak, compared with $\approx$3--5 times for AIA and STIX, reflecting the sharply-peaked compact kernel. While this sensitivity carries into the \hrieuv\ energy flux, it does not affect the qualitative result, as the \hrieuv\ flux remains an order of magnitude above the explosive threshold across the full range.

\end{appendix}

\end{document}